\documentclass[preprint,12pt,3p]{elsarticle}

\usepackage{setspace}
\usepackage{amssymb}
\usepackage{textcomp}
\usepackage{tikz}
\usepackage{etoolbox}
\usepackage{bm}
\usepackage{natbib}
\biboptions{sort&compress}
\usepackage{color}
\usepackage{MnSymbol}
\usepackage{amsmath}
\usepackage{algorithm}
\usepackage{algpseudocode}
\usepackage{mathtools}
\usepackage{caption}
\makeatletter
\def\BState{\State\hskip-\ALG@thistlm}
\makeatother

\definecolor{orange}{rgb}{1.0,0.4,0.0}
\definecolor{blueviolet}{rgb}{0.3,0,0.7}
\pdfoutput=1

\newcommand{\bphi}{
\boldsymbol{\phi}
}

\newrobustcmd*{\mytriangle}[1]{\tikz{\filldraw[draw=green,fill=#1] (0,0) --
(0.2cm,0) -- (0.1cm,0.2cm);}}

\newrobustcmd*{\mybarredtriangle}[1]{\tikz{\draw[draw=#1] (0,0) --
(0.2cm,0) -- (0.1cm,0.2cm) -- (0cm,0cm); \draw[draw=#1] (-0.1cm, 0.07cm) -- (0.3cm, 0.07cm)}}

\newrobustcmd*{\mybarredsquare}[1]{\tikz{\draw[draw=#1] (0,0)
rectangle (0.2cm,0.2cm); \draw[draw=#1] (-0.1cm, 0.1cm) -- (0.3cm, 0.1cm)}}

\newrobustcmd*{\mythickbarredsquare}[1]{\tikz{\draw[line width=0.4mm,draw=#1] (0,0)
rectangle (0.2cm,0.2cm); \draw[draw=#1] (-0.1cm, 0.1cm) -- (0.3cm, 0.1cm)}}

\newrobustcmd*{\mybarredcircle}[1]{\tikz{\draw[draw=#1] (0,0)
circle (0.1cm); \draw[draw=#1] (-0.2cm, 0.0cm) -- (0.2cm, 0.0cm)}}

\newrobustcmd*{\mythickbarredcircle}[1]{\tikz{\draw[line width=0.4mm,draw=#1] (0,0)
circle (0.1cm); \draw[draw=#1] (-0.2cm, 0.0cm) -- (0.2cm, 0.0cm)}}

\newrobustcmd*{\mydashedline}[1]{\tikz{\draw[draw=#1] (-0.2cm, 0.2cm) -- (-0.1cm, 0.2cm); \draw[draw=#1] (-0.0cm, 0.2cm) -- (0.1cm, 0.2cm)}}

\newrobustcmd*{\mybarredcross}[1]{\tikz{\draw[line width=0.4mm,draw=#1] (0,0) --
(0.2cm,0); \draw[line width=0.4mm,draw=#1] (0.1cm,-0.1cm) -- (0.1cm,0.1cm); \draw[draw=#1] (-0.1cm,0) -- (0.3cm,0);}}

\newrobustcmd*{\mythickline}[1]{\tikz{\draw[line width=0.4mm,draw=#1] (-0.15cm, 0.1cm) -- (0.15cm, 0.1cm);\draw[line width=0.4mm,draw=#1] (-0.0cm, 0.0cm);}}

\newrobustcmd*{\mythickdashedline}[1]{\tikz{\draw[line width=0.4mm,draw=#1] (-0.2, 0.1cm) -- (-0.1cm, 0.1cm); \draw[line width=0.4mm,draw=#1] (-0.0cm, 0.1cm) -- (0.1cm, 0.1cm); \draw[line width=0.4mm,draw=#1] (-0.0cm, 0.0cm);}}

\newrobustcmd*{\mycircle}[1]{\tikz{\draw[draw=#1] (0,0)
circle (0.1cm);}}

\newcommand{\norm}[1]{\left\lVert#1\right\rVert}

\DeclarePairedDelimiter\abs{\lvert}{\rvert}%

\journal{Journal of Computational Physics}

\begin{document}

\begin{frontmatter}

\title{Numerical convergence of the Lyapunov spectrum computed using low Mach number solvers}

\author[label1]{Malik Hassanaly\corref{cor1}}
\address[label1]{Department of Aerospace Engineering, University of Michigan, Ann Arbor, MI 48109, United States}

\cortext[cor1]{Corresponding author}
\ead{malik.hassanaly@gmail.com}

\author[label1]{Venkat Raman}
\ead{ramanvr@umich.edu}

\begin{abstract}

In the dynamical systems approach to describing turbulent or otherwise chaotic flows, an important quantity is the Lyapunov exponents and vectors that characterize the strange attractor of the flow. In particular, knowledge of the Lyapunov exponents and vectors will help identify perturbations that the system is most sensitive to, and quantify the dimension of the attractor. However, reliably computing these Lyapunov quantities requires robust numerical algorithms. While several perturbation-based techniques are available in literature, their application to commonly used turbulent flow solvers as well the numerical convergence properties have not been studied in detail. The goal of this work is two-fold: a) develop a robust algorithm for obtaining Lyapunov exponents and vectors for low-Mach based simulation of turbulent flows, b) quantify the spatial and temporal convergence properties of this algorithm using a series of progressively complex flow problems. In particular, a manufactured solutions approach is devised using the Orr-Sommerfeld (OS) perturbation theory to extract Lyapunov exponents and vectors from OS eigen solutions. While individual test cases show interesting results, overall the spatial convergence rates of Lyapunov exponents follow the truncation order for the discretization scheme. However, temporal convergence rates were found to be only a weak function of time step used. Additionally, for some configurations, the convergence properties are found to depend on the Lyapunov exponent itself. These results indicate that convergence properties do not follow universal rates, and require careful analysis for specific configurations considered. 
\end{abstract}

\begin{keyword}
Lyapunov spectrum, numerical convergence, attractor dimension, low Mach number solver
\end{keyword}

\end{frontmatter}


\newpage
\clearpage
\section{Introduction}

The use of dynamical systems (DS) approach for studying turbulent flows is regaining interest \cite{mohan2017scaling,qiqi_lss}, driven by the growth in computing power and the emergence of alternative uses for computational methods including the pursuit of sensitivity analysis \cite{qiqi} and optimization \cite{hicken2014pde}. More recently, the emergence of data-driven modeling, especially in the context of modal decomposition methods \cite{schmid2010jfm,williams2015data,brunton2017chaos,crom_kaiser} has re-ignited interest in the DS approach. Theoretical research in the past had established that turbulent flows fall in a class of chaotic flow problems that possess a strange attractor \cite{ruelle, temam}. Based on these ideas, the notion of inertial manifolds \cite{temam_manifold}, and approximate approaches for extracting these manifolds \cite{aim_temam, aim_kevrekidis} have been formulated. At the root of the DS approach is the ability to characterize the chaotic behavior of a flow by quantifying, for example, the dimension of its attractor. In this regard, the Lyapunov theory \cite{ruelle,grassberger2004measuring} is a useful mathematical tool.

In the practical use of Lyapunov theory, the growth of perturbations around a solution of a system of ordinary differential equations (ODE) is decomposed into a set of Lyapunov exponents (LEs) and vectors (LVs) \cite{benettin1,benettin2}. Altogether, they form the Lyapunov spectrum (LS). For a turbulent flow, only a finite subset of LEs are positive, and the dimension of the strange attractor of the flow can therefore be obtained using the Kaplan-Yorke estimate \cite{kaplan1979chaotic}. Beyond the original application of such techniques, LEs and LVs are beginning to be broadly applied, often in the context of uncertainty propagation. For instance, such perturbation techniques are used to determine growth of initial uncertainty in weather forecasting \cite{kalnay,kalnayBook}. In general, the focus is on determining the principal directions in phase-space along which perturbations can grow, leading to different formulations for such growth vectors \cite{wolfe}. More recently, such techniques have been extended to turbulent flows, where the adjoint solution of chaotic flow simulations are related to these LVs \cite{qiqi}. Due to the differing nature of these applications, there are numerous formulations of the LVs. In the classical method \cite{benettin1,benettin2} used to obtain the LEs, the growth of orthogonal perturbations is recorded. The resultant vectors are called Gram-Schmidt vectors (GSV), named after the orthogonalization procedure. Because the GSV are arbitrarily oriented, they are in general not directly used to analyze the chaoticity of a system. On the other hand, covariant Lyapunov vectors (CLV), which can be constructed from the GSV \cite{ginelli}, provide a description of the sub-spaces tangential to the attractor in which perturbations expand due to chaoticity \cite{inubushi}. In this work, the GSVs will be considered, and the focus is on the LEs rather than the vectors themselves. For this reason, the set of LEs will be referred to as the LS here. 

In the past, LEs and LVs have been calculated for a number of canonical turbulent flows. Keefe et al.~\cite{keefe} simulated a turbulent channel flow, albeit at low Reynolds numbers and with modest resolution that was tractable on machines of that time. They estimated the dimension of the strange attractor by determining a partial spectrum of LEs. LEs were also computed in order to quantify the predictability of chaotic systems \cite{metais-lesieur-predict}. Vastano et al.~\cite{vastano_moser} used the first finite-time LV to identify instability mechanisms. Inubushi et al.~\cite{inubushi} studied the growth of instability using multiple CLVs. Wang~\cite{qiqi} used the CLVs as a basis to compute the sensitivity of long-time averages in chaotic flows. Mohan et al.~\cite{mohan2017scaling} used the first LE as a measure of the chaoticity of the system, and studied its scaling as a function of Reynolds number for homogeneous isotropic turbulence (HIT). Taking into account the uncertainty of the sampling time necessary to obtain the LE, they established that the first LE grew at a higher rate than the Kolmogorov frequency. The focus of these studies has been mainly to demonstrate the applicability of the Lyapunov approach to study dynamics of perturbations.

In general, the LEs and LVs are computed using the procedure introduced by Benettin et al.~\cite{benettin1,benettin2}. While the algorithm is described in detail in Sec.~\ref{sec:algo} below, it is briefly outlined here to motivate the current work. For a dynamical system of dimension $\mathcal{D}$, there exists $\mathcal{D}$ LEs and an equal number of LVs, each with size $\mathcal{D} \times 1$. To compute the $m$ leading LEs, a set of $m+1$ simulations are conducted, where a baseline trajectory, and $m$ additional simulations with slightly perturbed initial conditions are used. These perturbations are obtained from $m$ initially orthonormal vectors. For ergodic systems, regardless of the initial perturbation, all the perturbations will eventually align along the leading LV (i.e., the LV corresponding to the fastest growing direction in the tangent space of the dynamical system). Since the perturbations grow exponentially, the perturbation vectors are periodically extracted, re-orthonormalized, and the simulations continued. The LEs are based on a time-average of the growth of perturbations. This procedure introduces many numerical issues. For instance, Keefe et al.~\cite{keefe} showed that refining the computational domain led to changes in the LE spectrum, even when the underlying simulations used pseudo-spectral, and hence highly accurate approaches. Further, the orthonormalization procedure to obtain the normalized Gram-Schmidt vectors is also subject to errors \cite{giraud2005loss}. Finally, the time-stepping procedure was found to introduce another error in the final LE spectrum, with the conclusion that a finer temporal discretization is needed to obtain converged solutions~\cite{keefe}. Recently, Fernandez et al.~\cite{fernandez} conducted a systematic study to determine the role of mesh size and order of numerical scheme on the Lyapunov exponents. While the study showed that LEs depend on the grid size and the numerical order, a clear convergence could not be established since the underlying forward solution itself was highly sensitive to these choices. Further, when perturbations are introduced at each re-orthonormalization step, there is a finite time taken by the system to reach the chaotic attractor, which introduces another factor to consider in the time-averaging used to estimate LEs.

In addition to the above issues, the focus of the current paper is in the use of certain special class of fluid solvers, typically employed for low Mach number flows \cite{harlow_stagg,pierce_thesis,desjardins-jcp,malik-openfoampaper}. These methods are applicable for simulating turbulence using both direct numerical simulation (DNS) and large eddy simulation (LES) approaches, specifically in the context of reacting and variable-density flows \cite{yihao-3-jet,alex-dlr,koodlr}. Central to these methods is the use of a Poisson equation to obtain a continuity-preserving velocity field at each time-step \cite{kimmoin-projection}. This step could be a source of numerical issues, especially when there is a need to ensure conservation of secondary quantities \cite{ham-les-complex-lowmach,hamiac-kecons,mahesh-jcp, malik-openfoampaper}. While some of the prior Lyapunov-theory studies have utilized low Mach number solvers, it has been mainly in the context of constant density periodic flows using spectral methods. On the other hand, Ref.~\cite{fernandez} used a fully compressible formulation, but with a discontinuous-Galerkin (DG) formulation of the spatial derivatives. 

With this background, this work is organized as follows: a) develop a computational procedure for computing a spectrum of LEs and LVs using a low Mach number variable density solver, b) use a progression of test cases to determine the relation between the numerical discretization errors associated with the spatial and temporal derivatives and the errors in the LE computation, and c) demonstrate this technique for canonical turbulent and reacting flows.

\section{Computation of LEs using low Mach number solvers}
\label{sec:algo}

With a dynamical systems perspective, the partial differential equations (PDE) governing the fluid flow and combustion are converted into a finite-dimensional set of ordinary differential equations (ODE). This is equivalent to assuming a sufficiently fine grid, and discretizing the equations. The resulting set of ODEs can be represented as
\begin{equation}
\frac{d\boldsymbol{\xi}}{dt} = \mathcal{F}(\boldsymbol{\xi});~~\boldsymbol{\xi}(t=0) = \boldsymbol{\xi}^0,
\label{eq:g}
\end{equation}
where $\boldsymbol{\xi}$ is the vector of variables, and $\mathcal{F}$ is the set of discretized governing equations. In terms of PDE, if a computational grid with $N$ control volumes/grid points is used, and $\boldsymbol{\xi}$ depends on $S$ fields, the length of the corresponding $\boldsymbol{\xi}$ vector will be $\mathcal{D}=S \times N$, which is the dimension of the state space for this system. Note here that the number of transported equations may be lower or equal to $S$. More details about the definition of the state vector are provided in Sec.~\ref{sec:defVector}. The state space is high-dimensional and will be described using a linear combination of spanning vectors. These vectors will be called state vectors or solution vectors in the rest of the paper.

The main objective is to determine the response of the flow to a perturbation. Using the aforementioned notations, one can consider a variation (perturbation) in state-space, given by $\delta \boldsymbol{\xi}$, which will evolve according to
\begin{equation}
\frac{d\delta\boldsymbol{\xi}}{dt} = \frac{\partial \mathcal{F}}{\partial \boldsymbol{\xi}}\delta \boldsymbol{\xi} = \mathcal{J}(\boldsymbol{\xi})\delta \boldsymbol{\xi} ;~~\delta{\boldsymbol{\xi}}(t=T_0) = \delta{\boldsymbol{\xi}}^0,   
\end{equation}
where $\delta{\boldsymbol{\xi}}^0$ is the initial perturbation at time $T_0$. The above equation can be integrated in time along with the governing equation (Eq.~\ref{eq:g}) to obtain the perturbation vector at time $T$
\begin{equation}
\delta \boldsymbol{\xi}(T) = \delta \boldsymbol{\xi}(T_0) \int_{T_0}^T \mathcal{J}(\boldsymbol{\xi}) dt,
\end{equation}
where the term inside the integral is independent of the perturbation. Based on this relation, an expansion rate can be expressed as
\begin{equation}
\Lambda = \frac{1}{T - T_0} \text{log} \frac{\norm{\delta \boldsymbol{\xi}(T)}}{\norm{\delta \boldsymbol{\xi}(T_0)}}.
\end{equation}
This quantity $\Lambda$  determines the rate at which an initial change to the flow field grows with time, between $T_0$ and $T$. If $\Lambda$ is negative, the system returns to the original state after a finite time (determined by the value of the expansion rate). In general, $\Lambda$ is a function of the local phase space (and therefore of the time) at which it is determined and $\Lambda$ can oscillate between negative and positive values. It is then possible to extract a mean value for the expansion rates as
\begin{equation}
\lambda \equiv \lim_{T \to \infty} \frac{1}{T} \int_0^T \Lambda(t) dt.
\end{equation}

$\Lambda$ is therefore the finite-time counter-part of $\lambda$, and will be called finite-time Lyapunov exponent (FTLE) in the rest of the paper. Here, FTLE \textbf{do not} denote Lyapunov exponents constructed using material derivatives \cite{apte-ftle,haller2011lagrangian}.
Since the original solution spans an $\mathcal{D}$-dimensional space, it is possible to obtain expansion rates for each of these dimensions by considering $\mathcal{D}$ such perturbations applied to the original system. The numerical procedure for obtaining these expansion rates became available in mid 1980s due to Benettin \cite{benettin1,benettin2}, which was then used to study a number of dynamical systems \cite{grappin,keefe,vastano_moser,kalnay}. The procedure consists in evolving an ensemble of simulations initially slightly perturbed from each other. The LEs are obtained by measuring the growth rates of the norm of the perturbations. In order to ensure that the growth rates computed characterize different subspaces of the phase space, the perturbations evolved are regularly orthogonalized. After each orthogonalization, they are aligned along the so-called Gram-Schmidt vectors (GSV), denoted as $\boldsymbol{\psi}$ in the rest of the text. Using Benettin's procedure, it is then possible to obtain the LS characterizing the dynamical response of the system. In general, the number of LVs and LEs is equal to the number of dimensions of the state-space. However, the procedure allows the selection of the first $m$ such vectors/exponents to be calculated, which ensures computational tractability. The set of LEs and LVs obtained is noted as

\[ \boldsymbol{\lambda} \equiv \{ \lambda_1, ..., \lambda_m\} \]
\[ \boldsymbol{\psi} \equiv \{ \psi_1, ..., \psi_m\} \]
 
Given this approach to obtaining the LS, the next step is to adapt the procedure for a low Mach number solver, which is discussed below. 
 
 \subsection{Definition of state vector}
 \label{sec:defVector}

 Establishing an analogy between a dynamical system and a low Mach solution approach for Navier-Stokes equations requires a precise knowledge of the numerical procedure. In general, the state vector should be defined such that its elements at the current time-step $n$, are necessary and sufficient to determine the state vector at the next time-step $n+1$. To see the importance of the definition of the state vector on the computed LEs, consider an $N$-dimensional dynamical system. In an ideal case, perturbations of the state vector can be expressed as
 \[
 \delta \boldsymbol{\xi}=\begin{bmatrix} \delta \xi_1^0 \\ \vdots \\ \delta \xi_N^0  \end{bmatrix},
 \]
 where the components of $\delta \boldsymbol{\xi}$ are necessary and sufficient to express any perturbation. The ``true" growth rate of a perturbation between the time-step $0$ and the time-step $n$ takes the form shown in Eq.~\ref{eq:truegrowth}.
 \begin{equation}
    \label{eq:truegrowth}
     \tau_{true} = \frac{|| \delta\boldsymbol{\xi}^n ||}{|| \delta\boldsymbol{\xi}^0 ||}.
 \end{equation}
 Now suppose that the definition of the perturbation vector is altered, for example by adding $N_{alt}$ components in the perturbation vector $\delta\boldsymbol{\xi}$. The added components are noted $\delta \boldsymbol{\zeta}$ which is a $N_{alt}$-dimensional vector. Using the L$_2$-norm, the ``altered" growth rate of the perturbation becomes:
 \begin{equation}
    \label{eq:alteredgrowth}
     \tau_{alt} = \sqrt{\frac{|| \delta\boldsymbol{\xi}^n ||^2 + || \delta\boldsymbol{\zeta}^n ||^2}{|| \delta\boldsymbol{\xi}^0 ||^2 + || \delta\boldsymbol{\zeta}^0 ||^2}}.
 \end{equation}
 A similar expression can be obtained if components are removed from the ``ideal" perturbation vector definition. Assuming that $|| \delta\boldsymbol{\zeta}^n || << || \delta\boldsymbol{\xi}^n ||$ and $|| \delta\boldsymbol{\zeta}^0 || << || \delta\boldsymbol{\xi}^0 ||$, one can obtain that $\tau_{alt} \approx \tau_{true}$. In many flow solvers, ghost cells are used to impose boundary conditions. The above conditions are satisfied even when these cells are not imposed in the state vector. In such cases, the condition $N_{alt} << N$, and the ghost cells are not expected to exhibit sensitivity to perturbations larger than other interior cells. However, these assumptions are in general not valid if an entire field is removed ad-hoc from the state vector. For instance, the entire determination cannot be based on only one component of velocity. In such cases, $N_{alt} \approx N$. While this may seem obvious, it is important to recognize that such choices are not self-evident in time-stepping algorithms that use outer-iterations within each time iteration to satisfy physical constraints. Beside the computation of the growth rates, the convergence properties of the algorithm described in Sec.~\ref{sec:algo} also depend on the orientation of the initial perturbation in the state space. With omission of variables, the orientations of the perturbation may be substantially modified, and may not guarantee convergence of the procedure. 
 
 \subsection{State vector for low Mach number solver}

As was previously explained in Sec.~\ref{sec:defVector}, defining the state-vector is a critical step that requires precise knowledge of the numerical procedure. The state vector is defined here for a variable density low Mach solver \cite{desjardins-jcp}. For the sake of completeness, the numerical procedure followed by the low Mach solver is first briefly described. Then, the difficulties associated with the state vector definition are discussed in relation to the numerical procedure.

\subsubsection{Numerical procedure of variable density low Mach number}

There exists many different version of low Mach solvers \cite{malik-openfoampaper,desjardins-jcp,shunn-MMS}, each requiring a slightly different definition of the state vector. Here, the general framework is described. Variable density low Mach number solvers provide solution to the following set of equations
 \begin{equation}
\label{eq:contNSmass}
\frac{\partial \rho}{\partial t} + \nabla \cdot (\rho \boldsymbol{u} ) = 0,
\end{equation}
\begin{equation}
\label{eq:contNSmom} 
\frac{\partial \rho {\boldsymbol{u}}}{\partial t} + \nabla \cdot (\rho {\boldsymbol{u}} {\boldsymbol{u}}) = - \nabla p + \nabla \cdot \boldsymbol{\overline{\sigma}},
\end{equation}
 where $\rho$ denotes the flow density, $\boldsymbol{u}$ the velocity vector, $p$ the mechanical pressure and $\overline{\boldsymbol{\sigma}}$ is the viscous stress tensor. In many reacting flow applications, energy release from these chemical reactions are assumed to affect only the internal energy, leading to a change in density and temperature of the fluid. For such applications, there are techniques that parameterize density changes based on a transported scalar (See, for instance, references within \cite{pitsch_arfm,ramanmalik_review}). Here, additional transport equations of scalar quantities are solved. A generic scalar transport equation can be written as
\begin{equation}
    \label{eq:scalarcont}
    \frac{\partial \rho \bphi}{\partial t} + \nabla \cdot (\rho \bphi \boldsymbol{u}) = \nabla \cdot (D \nabla \bphi) + \dot{\omega},
\end{equation} 
where $\bm{\phi}$ is the set of transported scalar, $D$ is the mass diffusivity and $\dot{\omega}$ is the scalar volumetric source term. In variable density applications, the thermochemical state (i.e, gas-phase composition including density and temperature) and transport parameters (diffusivity, viscosity etc.) are obtained as a function of $\bm{\phi}$. 

In the low Mach solver, density is not directly solved for, and is obtained from the scalar field \cite{pierce,pitsch-flamelet}. The right-hand-side of the scalar transport equation is written as $RHS_{\bm{\phi}}$. The momentum transport equations are advanced such that continuity is enforced at each time-step through a projection algorithm. One such algorithm (semi-implicit fractional time-step \cite{pierce_thesis}) is shown in Algo.~\ref{algo:lowMach} with Euler time discretization with a time-step $\Delta t$. Continuous derivatives are used when the details of the discretization are not relevant to the low Mach procedure. The momentum equation is advanced in two separate steps: first, an intermediate momentum (also called fractional momentum) $\rho \boldsymbol{u}^*$, is advanced using the best guess of the right-hand-side of the momentum equation $RHS_{\rho \boldsymbol{u}^*}$. Then the momentum equation is completed with a pressure term that enforces continuity. This algorithm follows a time-staggered procedure: at each step, $\boldsymbol{u}$ is advanced from time $n$ to time $n+1$ and $\boldsymbol{\phi}$ is advanced from $n+1/2$ to time $n+3/2$. Within each time-step, $N_{outer}$ outer-iterations (typically 3) are used to couple the momentum and scalar equations \cite{shunn-MMS,malik-openfoampaper}.

\begin{algorithm}
\caption{Variable density low Mach solver}\label{algo:lowMach}
\begin{algorithmic}[1]
\State At every grid point, initialize $\rho \boldsymbol{u}$,  $\boldsymbol{\phi}$
\For{$n=1,end$}

\State $\boldsymbol{u}^{n+1} = \boldsymbol{u}^{n}$, $\bm{\phi}^{n+3/2} = \bm{\phi}^{n+1/2}$

\For{$k=1,N_{outer}$}

\State $\bm{\phi}^{n+3/2} = \bm{\phi}^{n+1/2} + \Delta t RHS_{\bm{\phi}}$  
\State From $\bm{\phi}^{n+3/2}$ get $\rho^{n+3/2}$, $D^{n+3/2}$, ...
\State $\rho \boldsymbol{u}^* = \rho \boldsymbol{u}^n + \Delta t (RHS_{\rho \boldsymbol{u}^*}) $
\State Solve $\nabla^2 p = \frac{1}{\Delta t} (\frac{\partial \rho}{\partial t} + \nabla \cdot (\rho \boldsymbol{u}^*))$ 
\State $\rho \boldsymbol{u}^{n+1} = \rho \boldsymbol{u}^* - \Delta t \nabla p$

\EndFor

\EndFor
\end{algorithmic}
\end{algorithm}

\subsubsection{Variable density low Mach solver state vector}
\label{sec:stateVec}
At the very least, the state vector should contain all the transported variables which are independent from each other. Even if the variables are not defined at the same time locations, they are used together to define the next numerical step. Hence, the state vector should contain at least the fields $\{\rho \boldsymbol{u}, \boldsymbol{\phi}\}$. The rest of the set of discrete variables used in each time-step that needs to be included in the state vector depends on additional several subtle choices of the algorithm. In particular, the focus is here on the way the variables are used to form the guessed fractional momentum equation (See Algo.\ref{algo:lowMach}), and how the velocity field is obtained from the momentum field.

In the original formulation of the fractional time-step method \cite{kim-choi}, the fractional momentum equation contained a pressure gradient obtained from the previous time-step. In other terms, while the pressure term is solely dependent on the $\{ \rho \boldsymbol{u}^*, \boldsymbol{\phi}\}$, its contribution is accumulated over time. It is then necessary to include $p^{n}$ in the state vector in order to be able to write the relation between $\boldsymbol{\xi}^{n+1}$ and $\boldsymbol{\xi}^n$. Alternatively, the pressure can be left out of the state vector by setting it to zero at the beginning of each time-step. This only mildly affects the results of the procedure since outer-iterations help recover the correct guessed pressure. Here, the later alternative is chosen in order to minimize the number of variables in the state vector. 

The time-staggered representation used here creates variables at time stations $n+1/2$ and $n+3/2$ for the scalars, while momentum-related variables are advanced from $n$ to $n+1$. Hence, within each time step, a set of variables that are necessary for advancing the state vector is $\{ \rho \boldsymbol{u}^n, \boldsymbol{\phi}^{n+1/2}\}$. However, writing the momentum equation requires obtaining $\bm{u}$ at time-step $n$. In the current implementation, this is achieved by approximating $\rho^n$ as $\frac{\rho^{n+1/2}+\rho^{n-1/2}}{2}$. Hence, for complete determination of the state vector, it needs to be augmented to include information about $\rho^{n-1/2}$. Here, this variable is introduced as $\Delta \rho^n = \rho^{n+1/2} - \rho^{n-1/2}$, although other choices could have just easily been made. With the present formulation, the set of variables necessary and sufficient is $\boldsymbol{\xi}^n = \{ \rho \boldsymbol{u}^n, \boldsymbol{\phi}^{n+1/2}, \Delta \rho^n\}$

 \subsection{Modified algorithm for computing LEs}
 \label{sec:algorithm}

The numerical procedure originally formulated by Benettin et al.~\cite{benettin1,benettin2} for computing the LS involves multiple solutions of the dynamical system defined in Eq.~\ref{eq:g}. Here, this algorithm is modified to be suitable for the low Mach algorithm described above. First, the original algorithm is discussed, followed by the modification proposed here. For this purpose, Algo.~\ref{algo:benettin} shows the original algorithm \cite{benettin1,benettin2}. To compute the first $m$ LEs, multiple simulations are started from initial conditions obtained by adding a perturbation $\delta \boldsymbol{\xi}$ to a baseline solution $\boldsymbol{\xi}$. All simulations are advanced for $k_s$ time-steps after which the $m$ finite-time estimates are computed. For each simulation, the evolved perturbation corresponding to the difference between the solution of that realization and the baseline solution is obtained. These perturbations are orthogonalized using the Gram-Schmidt algorithm \cite{bjorck1994numerics,schmidt1908auflosung} which allows identification of the perturbation growth rates in $m$ directions orthogonal to each other. After each orthogonalization, $m$ FTLEs are obtained. The LEs are then obtained by repeating this procedure $N_s$ times and averaging the FTLEs over the $N_s$ cycles. In the following, $\mathcal{F}_k$ is the non-linear operator that advances a solution vector for $k$ time-steps of size $\Delta t$:

\[
  \mathcal{F}_k \colon \boldsymbol{\xi}(t=t_0) \mapsto \boldsymbol{\xi}(t=t_0+k\Delta t) ~~ \forall t_0
\]

\begin{algorithm}
\caption{Benettin's algorithm}\label{algo:benettin}
\begin{algorithmic}[1]
\State Randomly initialize $\{\delta \boldsymbol{\xi}_j\}_{1,m}$, orthogonal with norm $\norm{\delta \boldsymbol{\xi}_j} = \varepsilon$
\For{$i=1,$}
\For{$j=1,m$}
\State{$\delta\boldsymbol{\xi}{_j} = \mathcal{F}_{k_s}(\boldsymbol{\xi}+\delta \boldsymbol{\xi}_j) - \mathcal{F}_{k_s}(\boldsymbol{\xi})$}
\EndFor
\State Orthogonalize $[\delta\boldsymbol{\xi}_1, ..., \delta\boldsymbol{\xi}_m]$.
\For{$j=1,m$}
\State{$\Lambda_j = \frac{1}{k_s \Delta t} log(\frac{\norm{\delta\boldsymbol{\xi}_j}}{\varepsilon}) $}
\State{$\lambda_j = \lambda_j + \frac{1}{N_s} \Lambda_j$}
\EndFor
\State{Normalize $[\delta\boldsymbol{\xi}_1, ..., \delta\boldsymbol{\xi}_m]$, with norm $\varepsilon$}
\State{$\boldsymbol{\xi} = \mathcal{F}_{k_s}(\boldsymbol{\xi})$}
\EndFor
\end{algorithmic}
\end{algorithm}

When the perturbations are applied at the beginning of each cycle, there is no guarantee that the state vector for each simulation will be close to or lie on the attractor. If $\varepsilon$ is sufficiently but not too large, the system will temporarily lie within the basin of attraction. To reduce the distance from the attractor, $\varepsilon$ can be chosen to be small. While this approach works reasonably well to find positive LE, it creates numerical problems when computing negative LE (which are also required to obtain the Kaplan-Yorke dimension). For negative exponents, when the initial perturbation is too small, its magnitude may fall quickly below machine precision, leading to errors in the estimation of LEs \cite{malik_aiaa2017}.

To alleviate this issue, a modification to the LE algorithm is proposed.  While the perturbations are still orthogonalized with each other before applying it to the next set of simulations, their norms are not recorded immediately. Instead, the perturbations are evolved for $k_w$ time-steps, to provide a buffer time for the system to reach the attractor. An orthogonalization process takes place only to record the initial norm of orthogonal perturbation vectors, but the simulations are continued without any change in the state vector at this step. The simulations evolve, and the perturbation norms are then recorded a second and last time to compute the expansion rates of each perturbation. Currently, it is not feasible to \textit{a priori} determine the buffer time needed to reach the attractor. Instead a sensitivity analysis is used to determine the optimal buffer time. This modified procedure is detailed in Algo.~\ref{algo:modified}.

\begin{algorithm}
\caption{Modified algorithm}\label{algo:modified}
\begin{algorithmic}[1]
\State Randomly initialize $\{\delta \boldsymbol{\xi}_j\}_{1,m}$, orthogonal with norm $\norm{\delta \boldsymbol{\xi}_j} = \varepsilon$
\For{$i=1,$}
\For{$j=1,m$}
\State{$\delta\boldsymbol{\xi}{_j} = \mathcal{F}_{k_w}(\boldsymbol{\xi}+\delta \boldsymbol{\xi}_j) - \mathcal{F}_{k_w}(\boldsymbol{\xi})$}
\EndFor
\State Orthogonalize $[\delta\boldsymbol{\xi}_1, ..., \delta\boldsymbol{\xi}_m]$ and store the result as  $[\delta\boldsymbol{\zeta}_1, ..., \delta\boldsymbol{\zeta}_m]$.
\For {$j=1,m$}
\State{$\delta\boldsymbol{\xi}{_j} = \mathcal{F}_{k_s-k_w}(\mathcal{F}_{k_w}(\boldsymbol{\xi})+\delta\boldsymbol{\xi}_j) - \mathcal{F}_{k_s-k_w}(\mathcal{F}_{k_w}(\boldsymbol{\xi}))$}
\EndFor
\State Orthogonalize $[\delta\boldsymbol{\xi}_1, ..., \delta\boldsymbol{\xi}_m]$.
\For{$j=1,m$}
\State{$\Lambda_j = \frac{1}{(k_s-k_w)\Delta t} log(\frac{\norm{\delta\boldsymbol{\xi}_j}}{\norm{\delta\boldsymbol{\zeta}_j}}) $}
\State{$\lambda_j = \lambda_j + \frac{1}{N_s} \Lambda_j$}
\EndFor
\State{Normalize $[\delta\boldsymbol{\xi}_1, ..., \delta\boldsymbol{\xi}_m]$, with norm $\varepsilon$}
\State{$\boldsymbol{\xi} = \mathcal{F}_{k_s}(\boldsymbol{\xi})$}
\EndFor
\end{algorithmic}
\end{algorithm}

\subsection{Application of modified algorithm to turbulent partially-premixed flame}

\begin{figure}
\center
\includegraphics[width=0.4\textwidth,trim={0cm 0cm 0cm 0cm},clip]{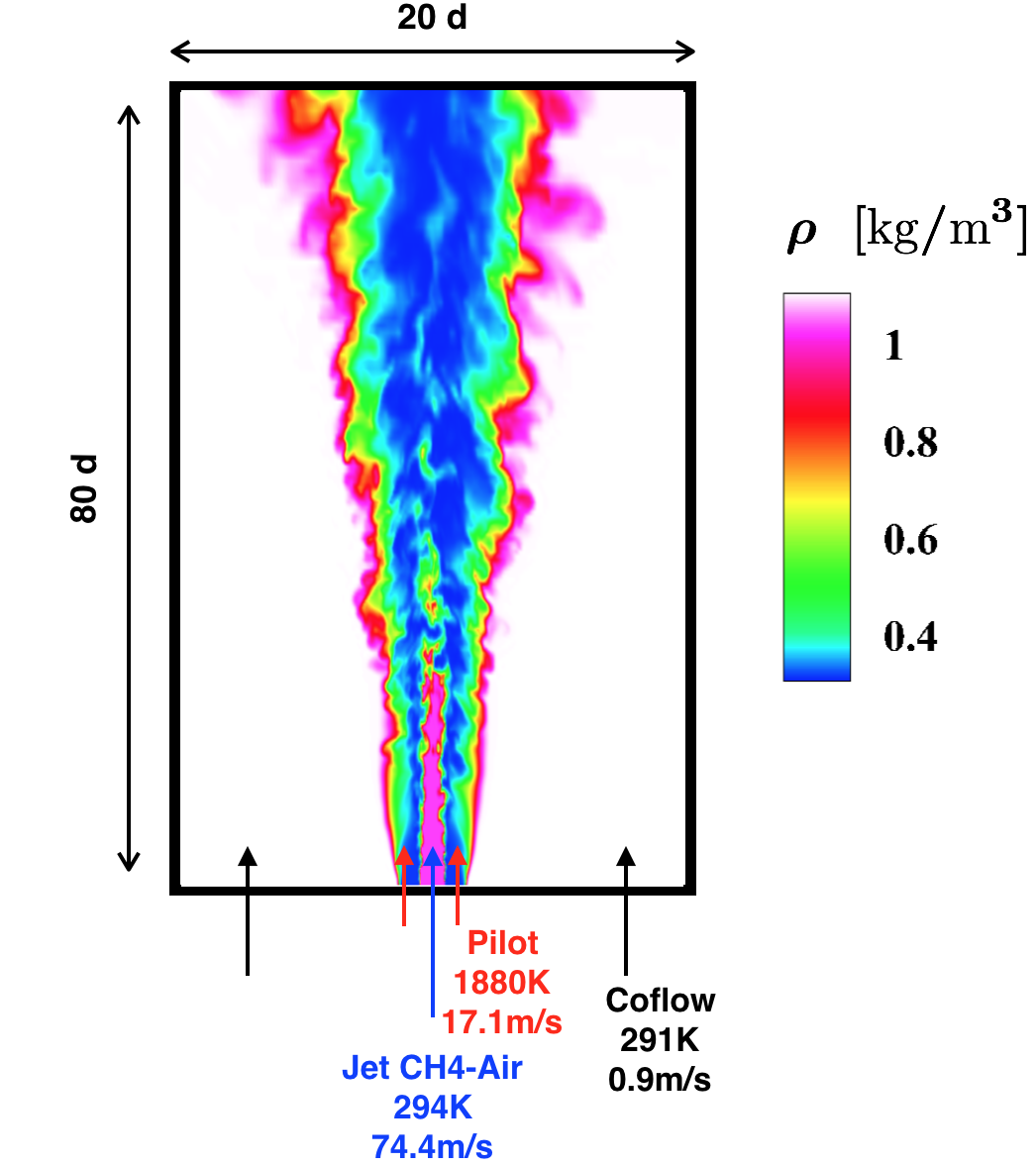}
\caption{Schematic of the Sandia E flame configuration simulated. The schematic is overlayed with an instantaneous contour of the density field.}
\label{fig:schematic}
\end{figure}

To recap, the modified algorithm records the expansion of perturbations after some buffer time used to ensure that the configuration has reached its attractor. A piloted turbulent flame test case is used to demonstrate the impact of the modified LE procedure (Sec.~\ref{sec:algorithm}). The test case replicates the widely studied E flame of the Sandia flame series \cite{barlow1998effects,tnf5,tnf6}. The flow configuration is shown in Fig.~\ref{fig:schematic}. This series of flames has been simulated with a wide range of combustion models over the past decade \cite{ramanpitsch-sandia1,pitschD,kempf,wang2008lagrangian,tnf4}. Here, a large eddy simulation approach is followed. Note that the goal of the current study is not to evaluate the models themselves, but the algorithm used to compute the Lyapunov exponents. In this study, the flamelet progress variable approach (FPVA) \cite{pierce,pierce-pof} is used. With these models, the variables that are solved include the filtered momentum vector ($\overline{\rho} \widetilde{\bm u}$), filtered mixture fraction ($\widetilde{Z}$), filtered progress variable ($\widetilde{C}$), as well as the sub-filter variance of mixture fraction ($\widetilde{Z'^2}$). The set of transport equations can be found in \cite{pierce}. The filtered density ($\overline{\rho}$) and transport properties are obtained from the look-up table. A cylindrical coordinate system with a grid composed of $256$ cells in the axial direction, $128$ in the radial direction and $32$ in the azimuthal direction is used.
 
The state vector is defined by following the reasoning of Sec.~\ref{sec:stateVec}. For a FPVA model, the scalars transported ($\boldsymbol{\phi}^{n+1/2}$) are the progress variable $\widetilde{C}$, mixture fraction $\widetilde{Z}$ and mixture fraction variance $\widetilde{Z'^2}$. As explained in Sec.~\ref{sec:stateVec}, three coordinates of velocity ($\overline{\rho}\boldsymbol{\widetilde{u}}^{n}$) should also be included in the state vector, along with the change in density between time-steps ($\Delta \overline{\rho}^n$). In total, the state vector comprises seven fields 
\[ \boldsymbol{\xi}^n = \{ \overline{\rho}\widetilde{u}^n, \overline{\rho}\widetilde{v}^n, \overline{\rho}\widetilde{w}^n, \widetilde{C}^{n+1/2}, \widetilde{Z}^{n+1/2}, \widetilde{Z'^2}^{n+1/2}, \Delta \overline{\rho}^n \}, \]
 \textit{i.e.} $\mathcal{D} = 7.3 \times 10^6$ variables.

The code used for this calculation implements physical and processor boundaries with ghost cells \cite{desjardins-jcp}. However, as explained in Sec.~\ref{sec:defVector}, only the interior points of the domain are considered in the state vector. For these calculations, the number of exponents computed is $m=30$. Following the algorithm described in Sec.~\ref{sec:algorithm}, $m+1$ independent simulations are commissioned, with one baseline simulations and $m$ simulations with initial conditions that are perturbations to the baseline condition. Here, the number of time-steps over which each expansion rate is computed is $k_s = 120$. The number of time-steps that are used as the buffer time is $k_w = 13$.  To investigate the effects of the modified algorithm, the norm of the perturbations corresponding to each simulation is measured at every time-step. The simulations are evolved over $N_s=119$ cycles, for a total simulation time of approximately three flow-through times ($2.38 \times 10^{-2}$~s). In this case, the initial perturbation norm is $\varepsilon=1 \times 10^{-3}$. 

\begin{figure}
\center
\includegraphics[width=0.4\textwidth,trim={0cm 0cm 0cm 0cm},clip]{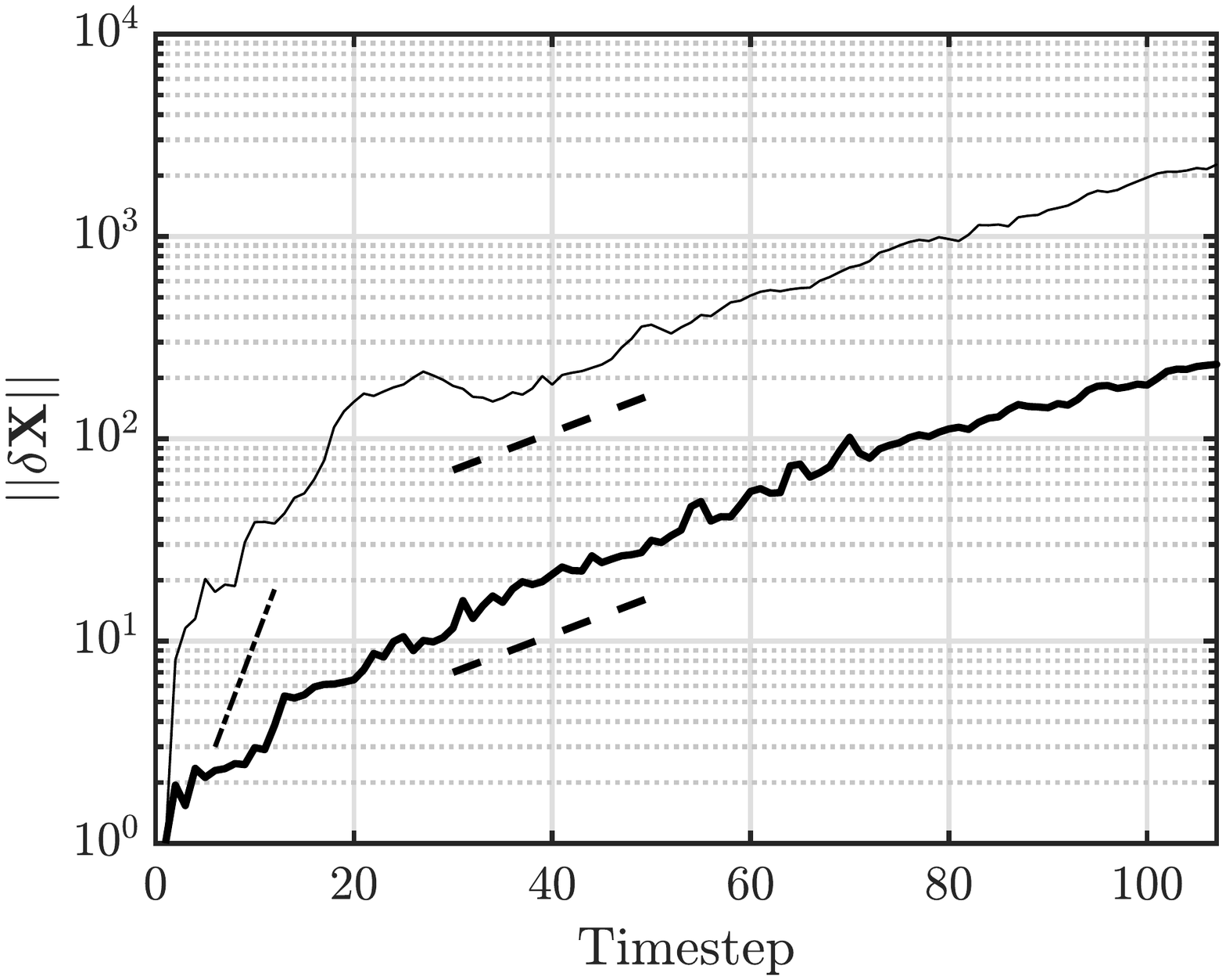}
\includegraphics[width=0.4\textwidth,trim={0cm 0cm 0cm 0cm},clip]{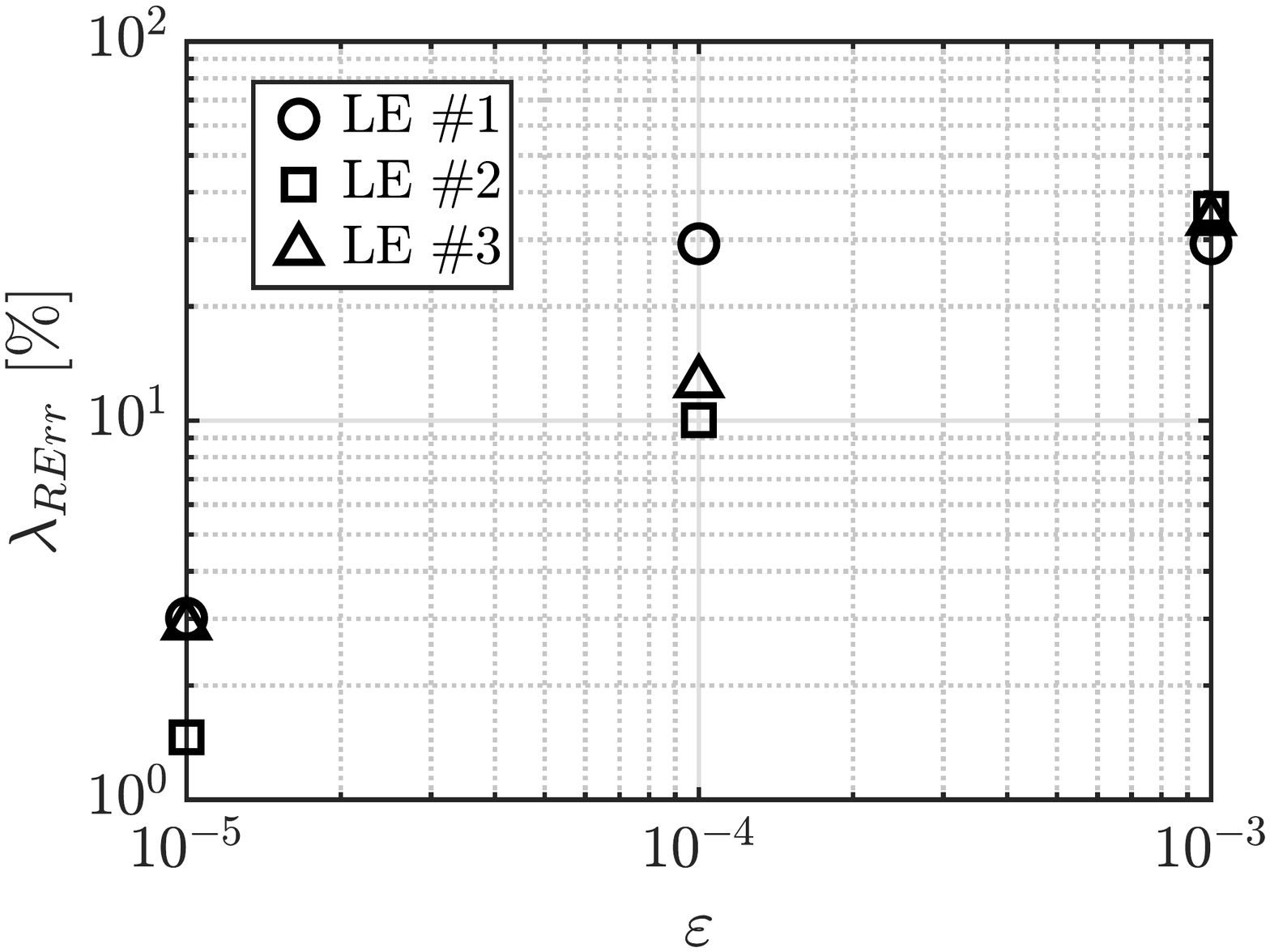}
\caption{Left: averaged time evolution of the $L_2$-norm of the 30 perturbations obtained for the classical algorithm (thin lines) and the modified algorithm (thick lines). The L$_2$-norm is normalized by its initial value ($10^{-3}$). The norm at every time instant is averaged over 119 realizations and the 30 LEs. Two different growth rates are shown with the original algorithm. They are outlined by the dashed and the dash-dotted line. Right: relative difference between the classical LE calculation and the modified LE calculation for the first three exponents. The error is plotted against the initial perturbation norm $\varepsilon$.}
\label{fig:modified}
\end{figure}

Figure~\ref{fig:modified} shows the evolution of the perturbation norm averaged over all the cycles. Note that given the Reynolds number of the Sandia E flame, the dimensionality of this problem is expected to be much larger than 30. Therefore the first 30 LEs are expected to be approximately constant compared to the spectrum of exponents. To show that the initial growth effect is present in general, the evolution of the perturbation norm was averaged over all the exponents. To provide a direct comparison between the modified and the classical algorithm, the first $k_s - k_w$ timesteps are plotted for the classical algorithm and the last $k_s - k_w$ are plotted for the modified algorithm. It is seen that when the original algorithm is used, the perturbation norm exhibits at first a higher slope, which marks the time taken for the system to return to its characteristic attractor. This growth rate is almost absent from the modified algorithm plot. It is further seen that the growth rates are nearly identical after this initial growth phase, meaning that using the classical algorithm would have led to an error in the LE values. 

To further confirm that this effect is only due to the perturbed solution not being on the attractor, the effect of the perturbation norm on the computed LE values is studied. Figure~\ref{fig:modified}~(right plot) shows the impact of the modified algorithm on the actual values of the LE for different initial perturbation norms $\varepsilon$. The LEs are computed by averaging the expansion rates across all the cycles of the algorithm. The classical algorithm records the perturbation norm expansion during $k_s$ timesteps for each cycle while the modified algorithm records it for the last $k_s - k_w$ timesteps. The error $\lambda_{RErr}$ plotted is given by
\begin{equation}
    \lambda_{RErr} = \frac{|\lambda_{modified} - \lambda_{classical}|}{|\lambda_{modified}|},
\end{equation}
where $\lambda_{classical}$ is the LE obtained with the classical algorithm and $\lambda_{modified}$ is the LE obtained with the modified algorithm. It can be seen that the impact of the modified algorithm decreases as the norm of the initial perturbation is reduced. First, this clearly indicates that the first $k_w$ timesteps introduce a different growth rate for the perturbations. It can be interpreted as reflecting the dynamics of the basin of attraction rather than the attractor itself. Second, the modified algorithm can be entirely replaced by using infinitesimally small perturbations. However, as mentioned above, this will lead to numerical issues for the negative exponents.

For the rest of the discussion below, the modified algorithm will be used with finite initial perturbation size. The studies presented below analyze the numerical convergence properties of the LE algorithm for a variety of canonical flows.

\section{Convergence of LS for laminar planar Couette flow}
\label{sec:couette}
The issue of numerical convergence of the LS was recently investigated by Fernandez et al.~\cite{fernandez}. While they were able to conjecture the convergence with spatial discretization, conclusive results were not obtained. It was noted that as spatial discretization was refined, the flow features were modified, thereby affecting the LEs. Since LEs are obtained from a time average of expansion rates, statistical errors could also contaminate LE computations. To overcome these issues, a laminar planar Couette flow is studied here. Because the flow is laminar, no uncertainty due to time averaging is introduced and since the attractor is a single point, the flow features are unaffected by spatial refinement. Furthermore, as noted by Keefe et al.~\cite{keefe}, this particular test case offers the possibility to estimate analytically the true set of LEs $\boldsymbol{\lambda}$ and its associated true set of LVs $\boldsymbol{\psi}$. The rigorous link between the stability analysis and the Lyapunov theory is detailed below, and the procedure to test the convergence of the LE is then described. 

Following Keefe et al.~\cite{keefe}, the domain size in the streamwise ($x$ direction), height ($y$ direction) and spanwise directions ($z$ direction) is [0,8 $\pi$] $\times$ [-1,1] $\times$ [0, 2$\pi$]. The domain is periodic in the $x$ and $z$ direction. At the conditions described in Ref.~\cite{keefe}, the viscosity is $\nu = \frac{1}{Re_{\tau}}$ and the streamwise velocity profile can be expressed as $U(y) =  \frac{Re_{\tau}}{2} (y^2 - 1)$, where $Re_{\tau}=34$. The analytical value of the pressure gradient can be obtained from the laminar channel flow equations. It is used for the forcing the laminar flow, as opposed to a more classical forcing term that would use the wall integral of the friction coefficient.

\subsection{Orr-Sommerfeld (OS) equations}

The OS equation has been formulated for several types of flows in the past \cite{grosch1978continuous}. The OS approach uses linearized governing equations to describe the evolution of small perturbations around the base flow. The main purpose here is to extract the stability properties of a given flow. The utility of the OS equation lies in its ability to describe a transient phenomenon with a reduced set of ODEs for which solutions can be obtained to a high-degree of precision \cite{orszag,dongarra}. Intuitively, the results of the OS analysis should be tightly linked to the results of the Lyapunov analysis. In Ref.~\cite{keefe}, it is in fact argued that the LEs should match the OS temporal eigenvalues up to a certain factor. In this section, this assertion is clarified further and used to determine convergence of LEs with grid refinement. 

For Couette flows, the OS equations have been formulated for two or three physical dimensions \cite{orr1907stability,sommerfeld1908beitrag}. Here, the three dimensional version is used. The full derivation of the three-dimensional OS equation can be found in Ref.~\cite[Chap 3.1]{schmidBook}. The perturbation to the flow field can be expressed in terms of wall normal velocity and vorticity perturbations as follows:
\begin{equation}
\label{eq:osdef}
\begin{split}
  \widehat{v}(x,y,z,t) = \widetilde{v}(y) e^{i(\alpha x + \beta z - c t)} ,
\\
\widehat{\eta}(x,y,z,t) = \widetilde{\eta}(y) e^{i(\alpha x + \beta z - c t)},
\end{split}
\end{equation}
where $\widehat{(.)}$ is the complex field whose real part satisfies the OS equation, $\widetilde{(.)}$ is the complex amplitude of $\widehat{(.)}$, and $\alpha$, $\beta$, $c$ are complex coefficients. The governing equations for the complex amplitudes $\widetilde{v}$ and $\widetilde{\eta}$ is then obtained as
\begin{equation} 
\label{eq:3Dos}
     \begin{split}  \Big\{  (-i c + i\alpha U)(\boldsymbol{D}^2 - (\alpha^2+\beta^2)) -i\alpha U'' - \frac{1}{Re_{\tau}}(\boldsymbol{D}^2 - (\alpha^2+\beta^2))^2 \Big\}\widetilde{v} = 0 \\ 
     \Big\{ (-i c + i\alpha U) - \frac{1}{Re_{\tau}}(\boldsymbol{D}^2 - (\alpha^2+\beta^2))^2 \Big\} \widetilde{\eta}  =  - i \beta U' \widetilde{v}, \end{split}
\end{equation}

where $\boldsymbol{D}$ is the derivation operator with respect to $y$.

The OS solution vectors are noted $\boldsymbol{\theta} \equiv \{ \widehat{v}, \widehat{\eta}  \}$ and are entirely defined by $\{ \widetilde{v}(y), \widetilde{\eta}(y), \alpha, \beta, c\}$. It should be noted that $\alpha$ and $\beta$ can hold only restricted values in order to satisfy constraints regarding periodicity in the streamwise and spanwise directions. As an illustration, let $\alpha = \alpha_1 + i \alpha_2$, where $\alpha_1$ and $\alpha_2$ are real numbers. Since the configuration is periodic in the $x$ direction, with spatial period $L_x$, and $\alpha$ is uniform and constant, 
\[ \forall~y,~t, ~~\widehat{v}(x=0,y,z=0,t) =   \widehat{v}(x=L_x,y,z=0,t). \]
Consider a $y$ coordinate for which the amplitude $\widetilde{v}(y) \neq 0$, then using Eq.~\ref{eq:osdef}
\[ \forall~t,  ~~e^{-i c t} = e^{-\alpha_2 L_x} e^{i (\alpha_1 L_x - c t)}. \]
Therefore $\alpha_2$ is necessarily zero and the set of realizable values of $\alpha_1$ can be restricted to $\alpha_1 = \alpha = n \frac{2 \pi}{L_x}$, where $n$ is an integer. Reproducing the same analysis in the spanwise direction, it can be shown that $\beta = m \frac{2 \pi}{L_z}$ where $m$ is an integer, and $L_z$ is the domain size in the spanwise direction.

\subsection{Link between Lyapunov exponents and OS eigenvalues}
\label{sec:link}

Consider a single LV $\psi$.  Since $\psi$ and the real part of the OS eigen solutions satisfy the same linear perturbation equation, the LV can be expressed as a linear combination of OS eigen solutions. More formally, it can be written as 
\begin{equation}
  \psi = \sum_j \alpha_j \operatorname{Re}(\boldsymbol{\theta}_j),
\end{equation}
where  $\alpha_j$ is a real coefficient. Note that the summation is over infinite set of solutions. Under the assumption that all the OS eigenvalues $c$ are different, any initial perturbation to the flow will converge to the eigen solution with the lowest decay rate. $\psi$ will converge to the OS solution with the amplitude that decays the least. If this perturbation could be matched to an eigen solution with characteristic wavenumbers $\alpha$ and $\beta$, then the Lyapunov exponent $\lambda$ can be related to the corresponding eigenvalue $c$ as 
\begin{equation}
    \lambda = \operatorname{Im}(c).
    \label{eq:c_le}
\end{equation}

Keefe et al.\cite{keefe} used this equivalence between OS solutions and LVs, but did not explicitly derive the relations shown in this section. The above derivation clarifies the equivalence invoked in Ref.~\cite{keefe}.

\subsection{Convergence of LE}

\subsubsection{Convergence procedure}
\label{sec:convProc}

The equivalence between OS solutions and LVs was used by Keefe et al.~\cite{keefe} to verify LS solutions. However, this approach requires that for each LV, the corresponding streamwise ($\alpha$) and spanwise wavenumbers ($\beta$) be manually identified from the computed LV. Here instead, one takes advantage of the knowledge of the full solution of the OS eigenvalue problem. A main observation is that when the perturbation for the full system is introduced based on the eigen solution for a given $\{\alpha,\beta\}$, the evolution in time retains this vector direction. As a result, the decay rate of the perturbation can be directly obtained and compared with the corresponding OS eigenvalue. 

The full procedure is as follows:
\begin{itemize}
    \item Chose a set of wavenumbers $\{\alpha,\beta\}$, and obtain the eigen solution to the OS problem. This will provide an eigen value that characterizes the Lyapunov exponent as shown in Eq.~\ref{eq:c_le}.
    \item Using the definition of $\eta$ and mass conservation equations, obtain $\widetilde{u(y)}$ and $\widetilde{w(y)}$. Together with $\widetilde{v(y)}$, a full vector of perturbations of dimension equal to that of the system can obtained.
    \item Evolve the perturbed simulation in time and obtain the decay rate of the perturbations. 
\end{itemize}
The LE convergence is evaluated in terms of the absolute error of the LE defined as 
\begin{equation}
    \lambda_{err} = |\lambda_{computed} - \lambda_{true}|,
\end{equation}
where $\lambda_{err}$ is the absolute error done for the LE estimation, $\lambda_{computed}$ is the computed LE, and $\lambda_{true}$ is the exact LE. Using the absolute error of the LE offers the advantage of evaluating the convergence of LE close to zero.

Note that the conventional approach of using random initial perturbations has some notable disadvantages compared to this approach. First, a long transient time will be needed before the solution converges to a particular LV. Second, it is not possible to isolate a single eigen solution but to compute multiple LVs at the same time, and match these with OS solutions in a postprocessing step. Such matching might itself be problematic if there are large numerical errors due to the discretized set of equations used. As a result, the above procedure is equivalent to a manufactured solutions approach for verifying an algorithm.

\subsubsection{Temporal convergence}

\begin{figure}
\center
\includegraphics[width=0.45\textwidth,trim={0cm 0cm 0cm 0cm},clip]{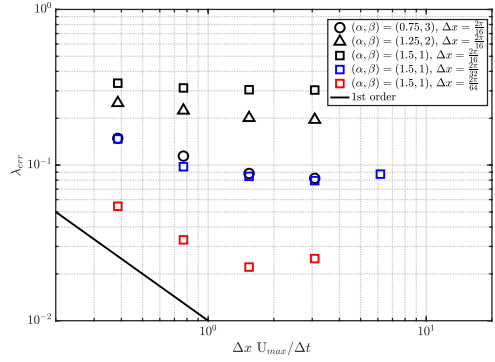}
\caption{Convergence of LE as a function of the time step. The exponent error is the difference between the observed LE and the Orr-Sommerfeld eigenvalue.}
\label{fig:convTS}
\end{figure}

For the temporal convergence tests, three different sets of parameters $\alpha$  and $\beta$ are considered: $(\{ 0.75, 3 \}, \{1.25, 2\}, \{1.5,1\})$. For all the parameters, calculations are conducted with a $16 \times 32 \times 16$ grid. In order to evaluate the relative impact of spatial and temporal discretization, a $32 \times 64 \times 32$ grid and a $64 \times 128 \times 64$ grid are additionally used with $(\alpha,\beta)=(1.5,1)$.  Figure~\ref{fig:convTS} shows the impact of the time discretization on the numerical estimate of the LE. One can observe a first order convergence of the LEs with respect to the time discretization for large timesteps. However this convergence is quickly stopped and plateaus for a convective CFL number larger than 1. The level of accuracy at which the convergence stops depends on the spatial discretization and will be investigated in greater details in the next section. The minimal effect of the timestep on the convergence of the error at small CFL numbers is consistent with the analysis Fernandez et al.~\cite{fernandez}, who used a compressible flow solver for an airfoil geometry and an acoustic CFL number ranging from 0.2 to 0.0125. Since the LEs are precisely determined here, there are no statistical errors associated with this evaluation. 

\subsubsection{Spatial convergence}

\begin{figure}
\center
\includegraphics[width=0.45\textwidth,trim={0cm 0cm 0cm 0cm},clip]{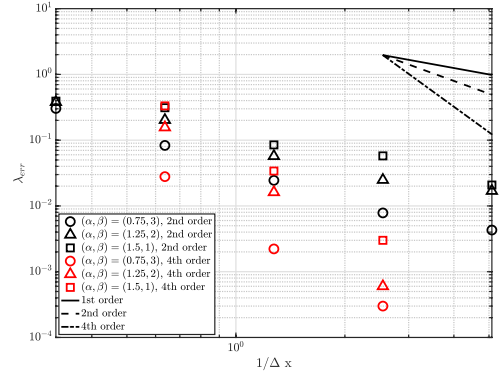}
\caption{Convergence of LE with spatial discretization. The exponent error is the different between the observed LE and the Orr-Sommerfeld eigenvalue.}
\label{fig:convSpace}
\end{figure}

The effect of grid spacing on LE calculation can be determined similar to the temporal convergence tests. Here, spatial convergence for different discretization truncation orders is studied. Starting from a grid of $8 \times 16 \times 8$, the grid spacing is progressively decreased by a factor of 2 each time to obtain the results shown in Fig.~\ref{fig:convSpace}. In order to clearly distinguish between the impact of temporal discretization and spatial discretization, the timestep is held fixed. For the $2^{nd}$ order case, the time-step is set to $7.5$ms; for the $4^{th}$ order case it is set to $1.625$ms. At these timesteps, the temporal errors do not dominate the errors in the approximation of LE. It is seen that the LE convergence rate is directly related to the order of numerical scheme used, with the slope of the error nearly matching the order of the scheme. To the authors' knowledge, this is the first such result for convergence for LEs. 

\subsubsection{LV-dependent error}

\begin{figure}
\center
\includegraphics[width=0.45\textwidth,trim={0cm 0cm 0cm 0cm},clip]{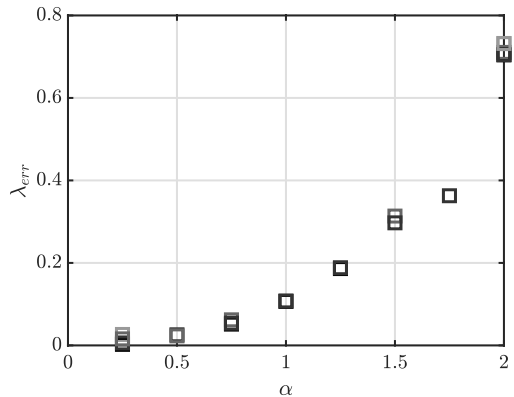}
\caption{Absolute error between the LE and the Orr-Sommerfeld eigenvalue plotted against the streamwise wavenumber of the Orr-Sommerfeld eigenvector. Darker colors indicate smaller spanwise wavenumber and lighter color indicate larger spanwise wavenumber (ranging from 0 to 3).}
\label{fig:ErrStream}
\end{figure}

Although the the LEs converge with spatial discretization, the absolute errors for individual LEs span a wide range that follows an interesting pattern. Figure~\ref{fig:ErrStream} shows the exponent error for LEs (defined similar to Fig.~\ref{fig:convTS} and Fig.~\ref{fig:convSpace}) as a function of the streamwise wavenumber for a single computational grid $16 \times 32 \times 16$. It is seen that there exists a strong dependence on the wavenumber, which is comparable to dispersion errors arising from Taylor's series expansion based discretization schemes \cite{kravmoin-leserrors}.  This trend can also be observed in Fig.~\ref{fig:convSpace}. This results was verified using two independent procedures. First, similar to Keefe et al. \cite{keefe}, the LS was computed using Benettin's algorithm. The streamwise and spanwise wavenumbers of the LV were individually extracted allowing to compare the LE with the eigenvalues obtained from the OS analysis. Second, similar to the procedure used in Sec.~\ref{sec:convProc}, initial conditions aligned with the eigenvectors obtained from the OS analysis were imposed and the decay rate of the perturbations were measured. Both procedure showed a similar dependence with respect to the streamwise wavenumber.

\section{Numerical convergence of Lyapunov spectrum for unsteady problems}
\label{sec:transient}
Although the laminar Couette flow problem is an interesting unsteady validation case for the LS algorithms, Lyapunov theory is more useful in the context of turbulent flow with strange and multidimensional attractors. In particular, an accurate estimation of the dimension is a first step in obtaining a high fidelity envelope for the attractor. In this section, a suite of test cases is used to determine the convergence properties of the LS for such transient problems.

\subsection{Kuramoto-Sivashinsky equation (KS) equation}
\label{sec:ks}
The KS equation represents a canonical configuration with turbulence-like chaotic behavior \cite{kuramoto,sivashinsky}. Due to its one-dimensional formulation, it is a computationally efficient model for studying chaotic behavior using the Lyapunov approach. With unity diffusion coefficient, the KS equation is written as
\begin{equation}
\frac{\partial u}{\partial t} + \nabla^4 u
 + \nabla^2 u + \nabla u^2 = 0, 
 \end{equation}
where $u(x,t)$ is the solution of the KS equation, defined on $x=[0, 16 \pi]$ and $t=[0,400]$, with $u(x,0) = \text{cos}(x/16)*(1+\text{sin}(x/16))$. For this study, a spectral approach is used. As will be discussed below, this method is modified to reflect some of the properties of the low Mach number solver. The PDE is using an exponential time differencing Runge-Kutta method (ETDRK4) \cite{etdrk4,cox2002exponential} with time-step fixed at $0.05$s. The non-linear term is treated with a pseudo spectral approach and $2/3$ dealiasing.

\begin{figure}
\center
\includegraphics[width=0.45\textwidth,trim={0cm 0cm 0cm 0cm},clip]{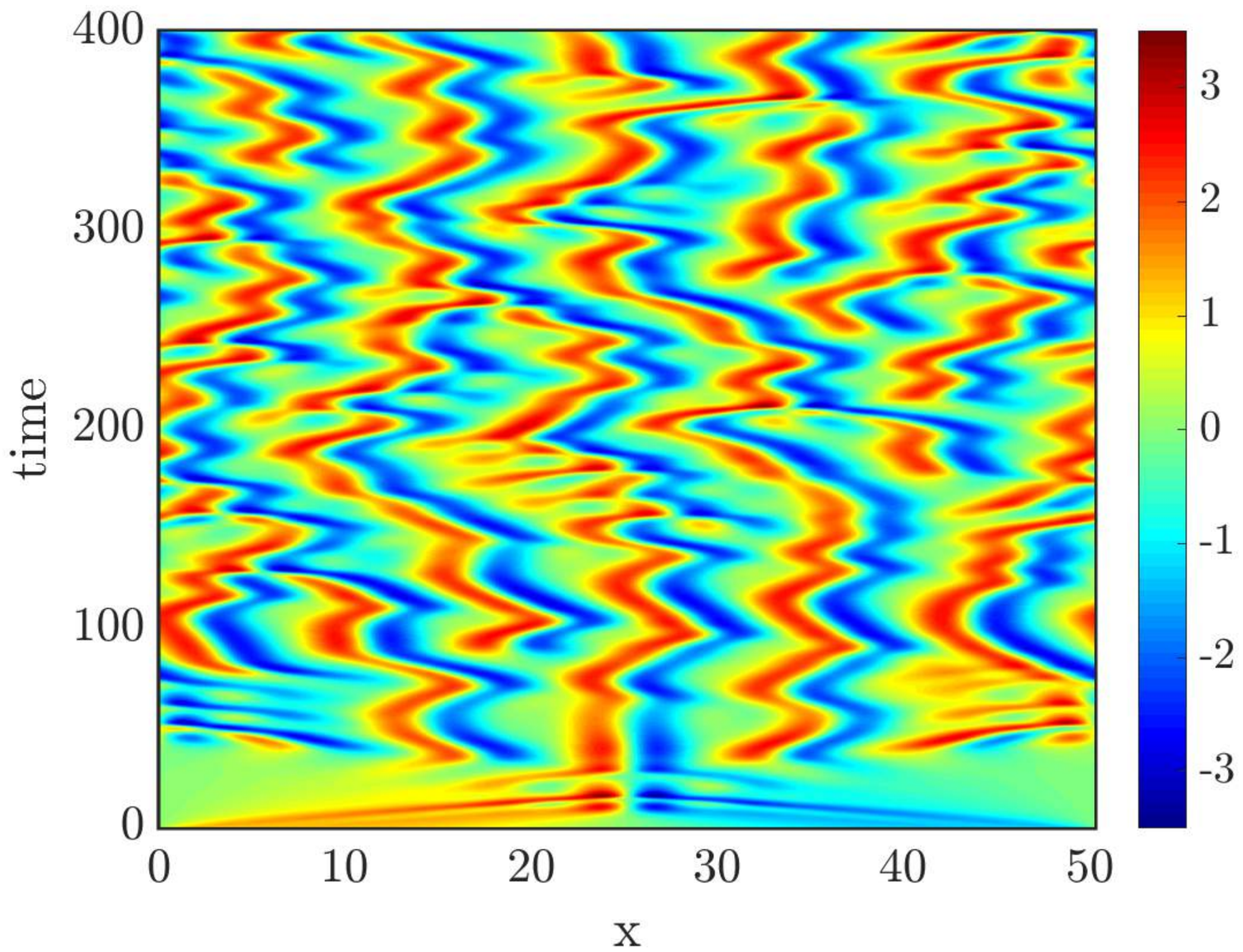}
\includegraphics[width=0.45\textwidth,trim={0cm 0cm 0cm 0cm},clip]{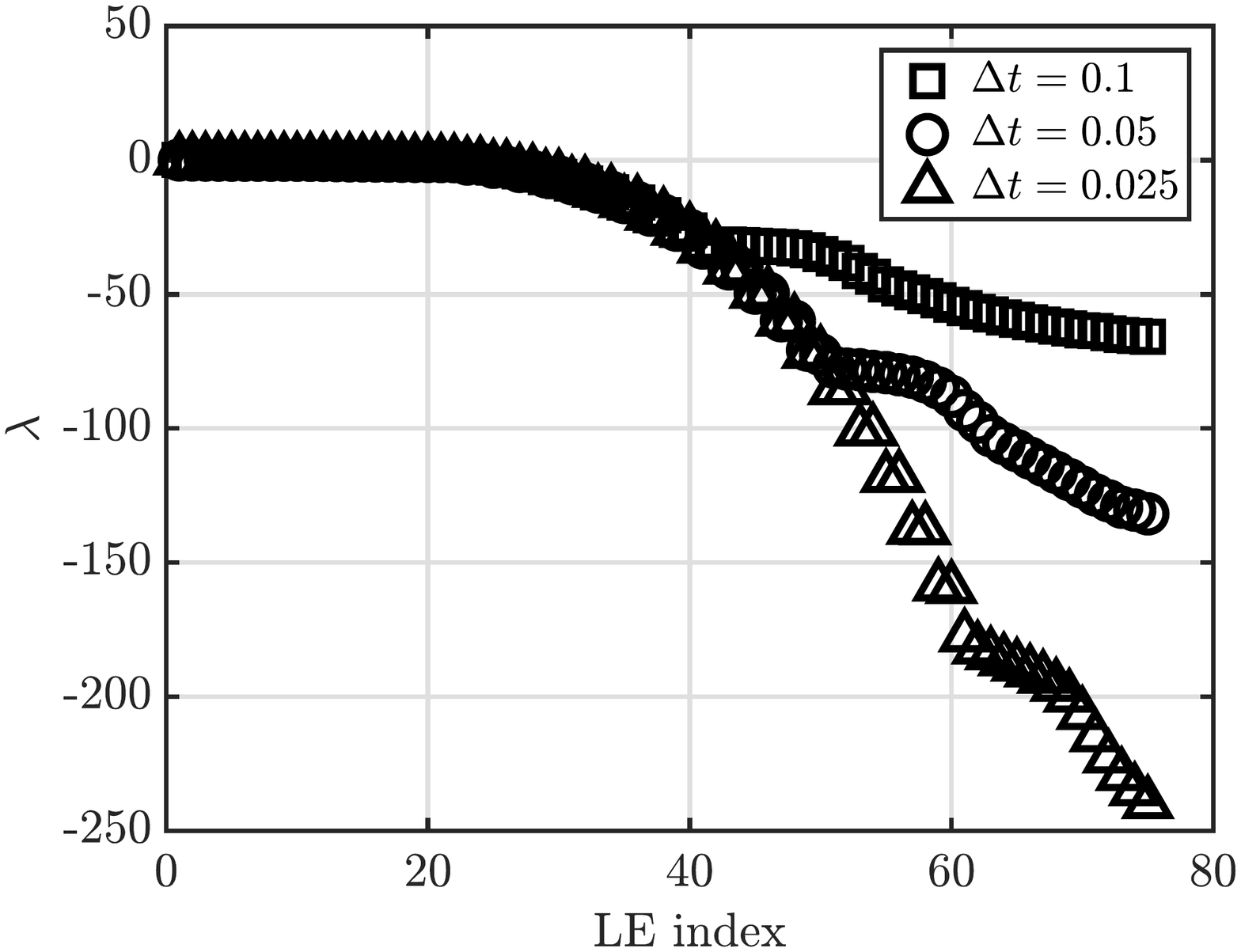}
\caption{Left: time evolution of the solution to the KS equation considered here. Right: first 75 LE obtained using spectral method with 500 modes.}
\label{fig:ksplots}
\end{figure}

Figure~\ref{fig:ksplots} shows the resulting space-time plots for a spatial discretization using 500 grid points. The first 75 exponents of the LS are computed in the statistically stationary region after $t=40$s (shown in Fig.~\ref{fig:ksplots}). The L$_2$-norm of the perturbations applied was set to $0.1$. The perturbation norms are recorded after waiting one time step, and are averaged over three time-steps before being orthogonalized. As can be seen, in Fig.~\ref{fig:ksplots}, the timestep size has a substantial impact on the last exponent of the spectrum. Because of the particular time integration scheme used, the temporal discretization dependence of the LS is not investigated here. Instead, Fig.~\ref{fig:ksplots} indicates that a timestep $0.05$s is sufficient to study the dependence of the first 45 exponents with the time discretization. Note that the initial perturbation norm was chosen such that the fastest decaying perturbation does not go below machine precision. The KS equation for this set of conditions exhibits a strange attractor for 5 strictly positive Lyapunov exponents. 

The impact of spatial discretization is assessed in this problem in two ways: a) by changing the number of grid points from 76 to 500 progressively, and b) by altering the truncation order of the spectral solver using a modified wavenumber approach \cite{kravmoin-leserrors}. Here, modified wavenumbers for second and fourth order central difference schemes are used in addition to the spectral derivatives. In order to be consistent with the low Mach number solver used in the rest of the study, the modified wave number is based on a staggered spatial discretization \cite{desjardins-jcp}. Since the FTLEs are time varying, the statistically averaged global LEs are evaluated using multiple re-orthonormalization steps. To assess the sampling uncertainty resulting from the use of a finite number of such steps, the method outlined in Ref.~\cite{mohan2017scaling,uncertaintyDNS_oliver} is used.  In all the plots, the LS obtained with the spectral method using 500 points is considered to be the ``true" LS.

\begin{figure}
\center
\includegraphics[width=0.32\textwidth,trim={0cm 0cm 0cm 0cm},clip]{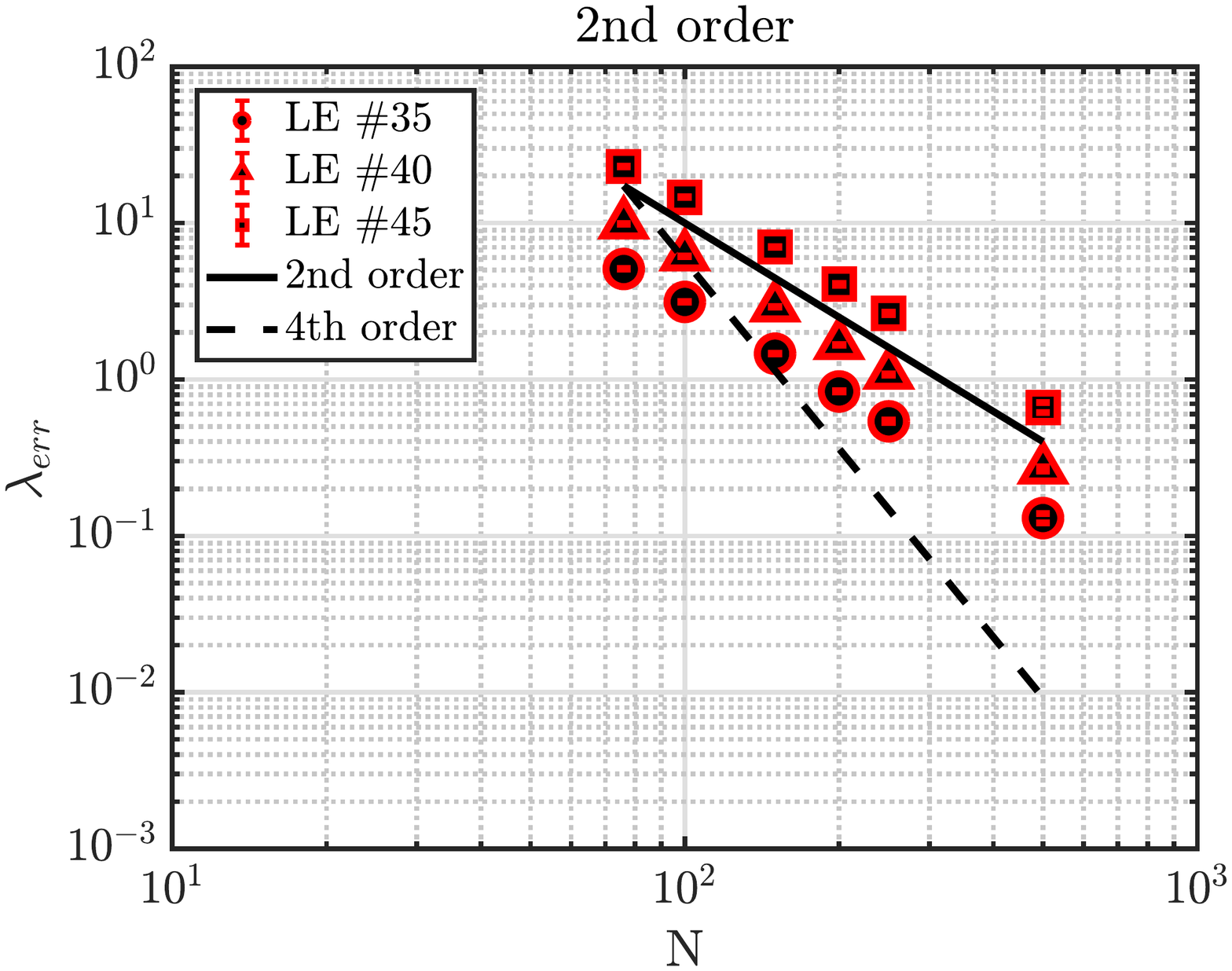}
\includegraphics[width=0.32\textwidth,trim={0cm 0cm 0cm 0cm},clip]{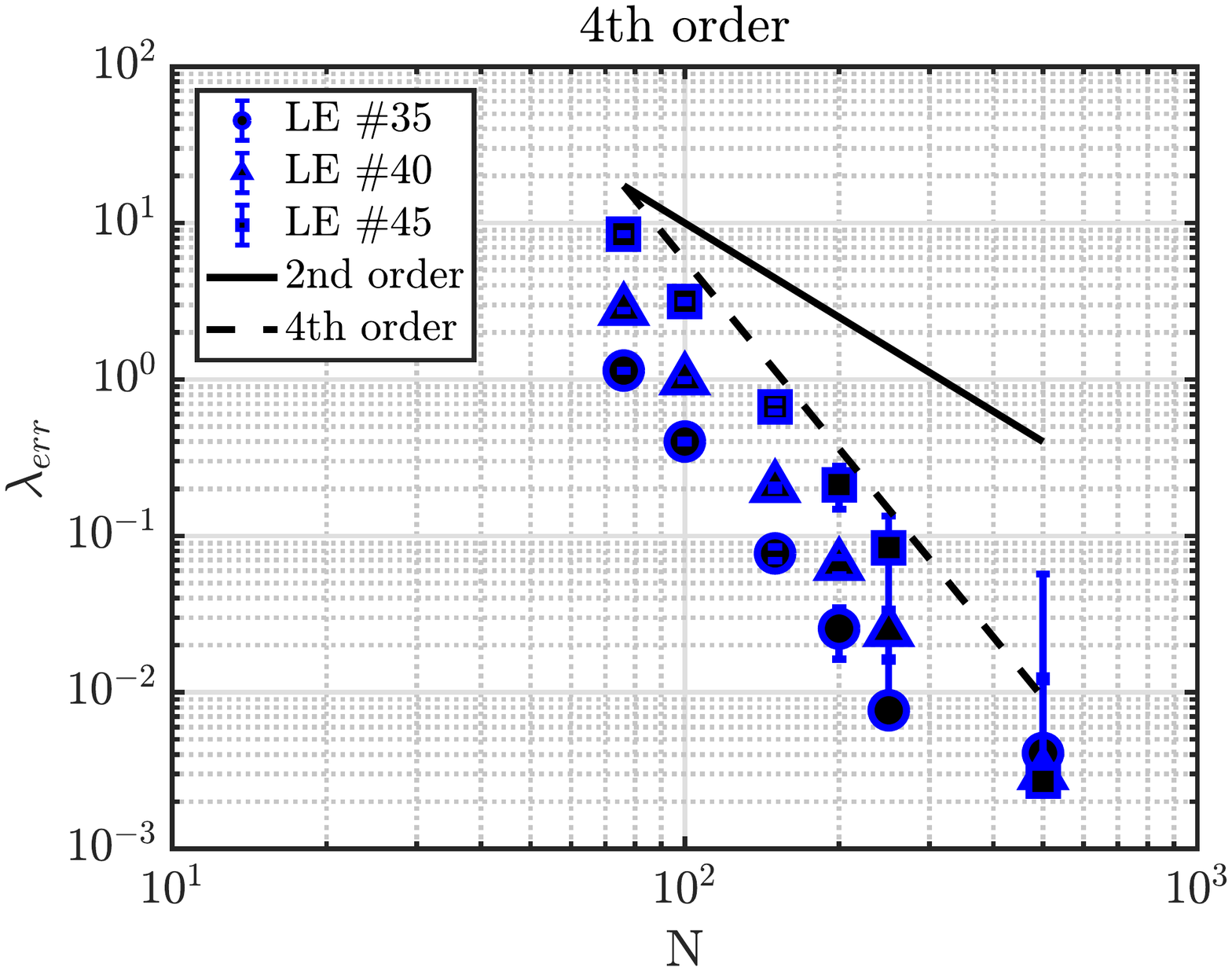}
\includegraphics[width=0.32\textwidth,trim={0cm 0cm 0cm 0cm},clip]{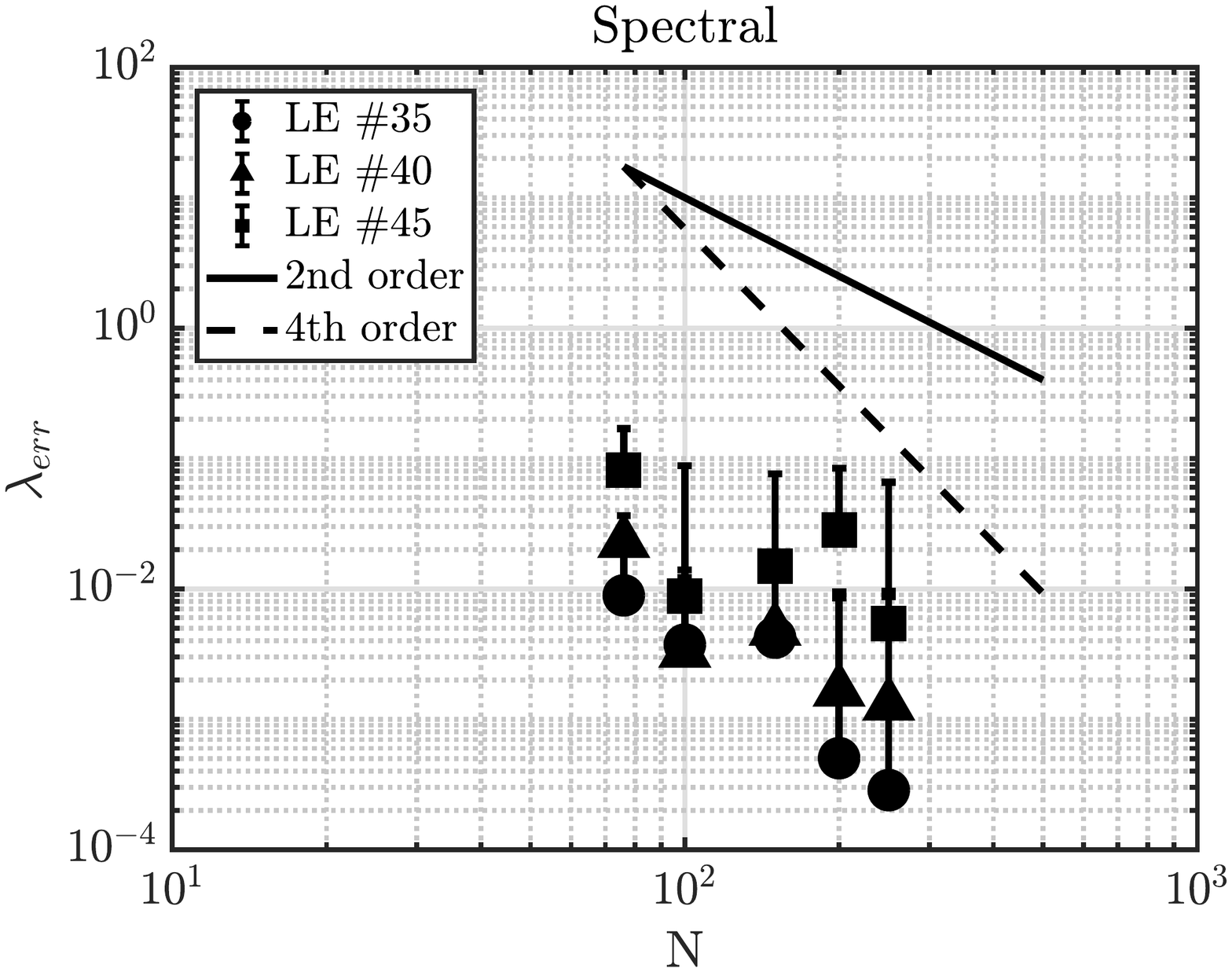}
\caption{Convergence of the $35^{th}$, $40^{th}$, $45^{th}$ LE with discretization level. Left: $2^{nd}$ order. Middle: $4^{th}$ order. Right: spectral method.}
\label{fig:KSgood}
\end{figure}

\begin{figure}
\center
\includegraphics[width=0.32\textwidth,trim={0cm 0cm 0cm 0cm},clip]{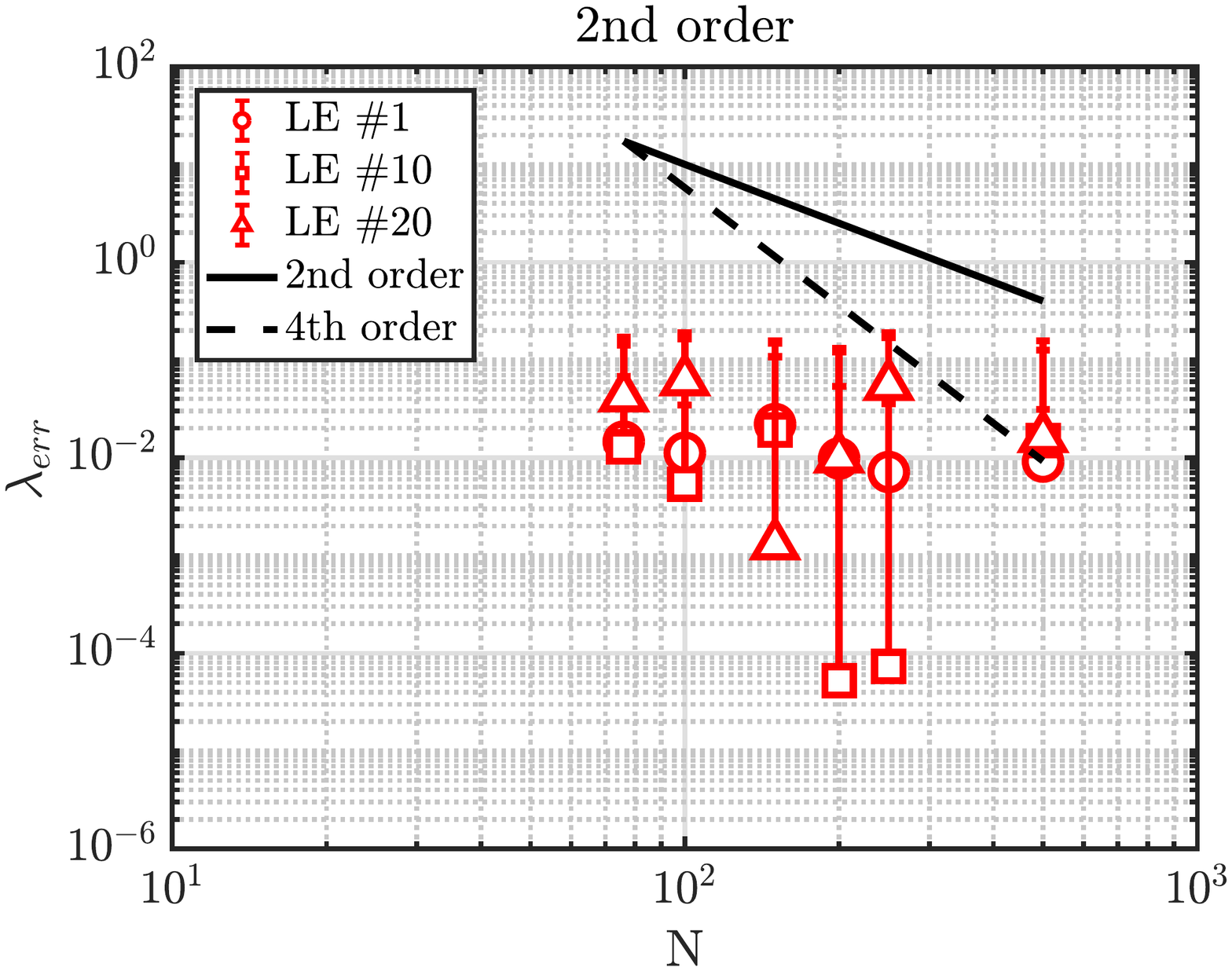}
\includegraphics[width=0.32\textwidth,trim={0cm 0cm 0cm 0cm},clip]{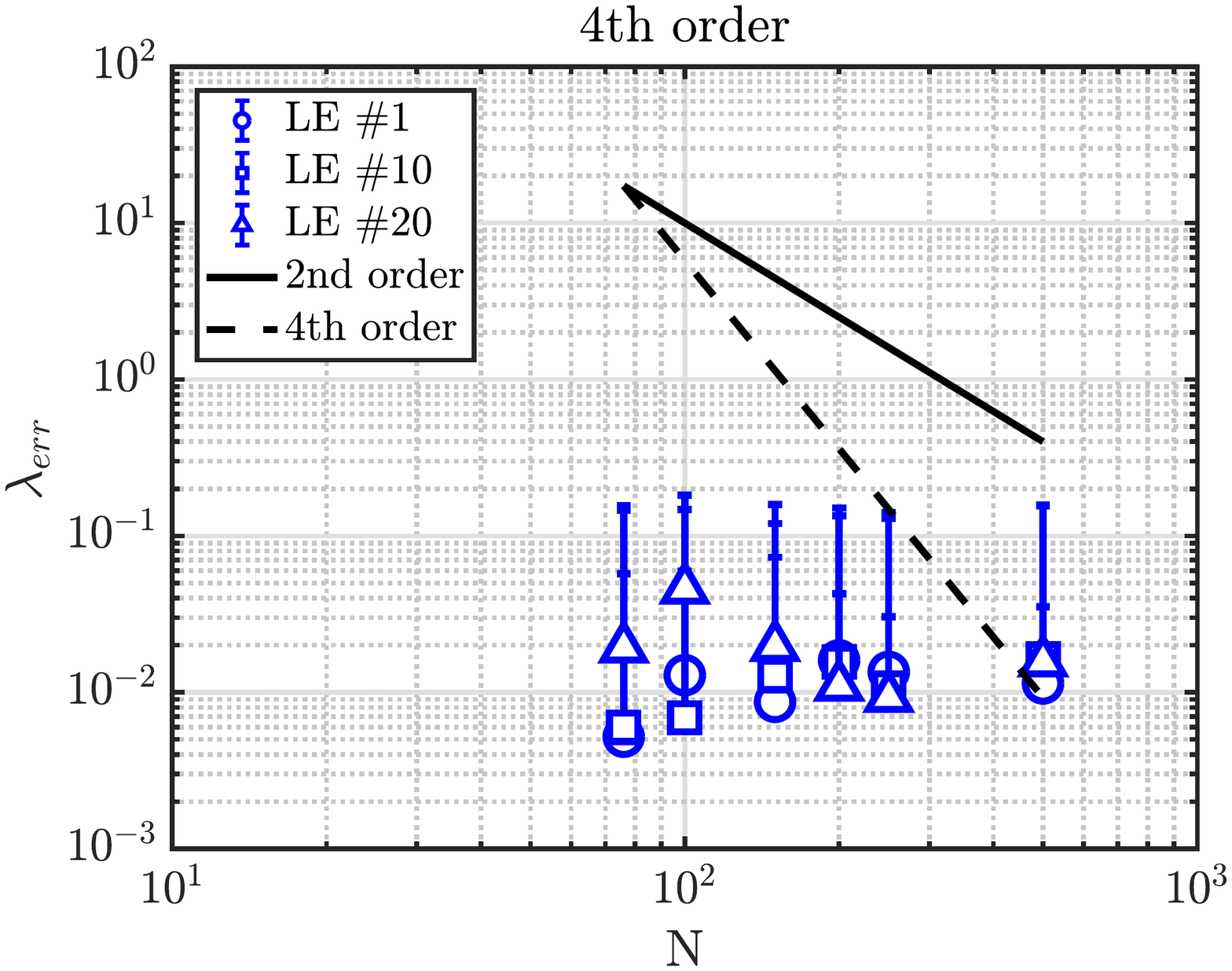}
\includegraphics[width=0.32\textwidth,trim={0cm 0cm 0cm 0cm},clip]{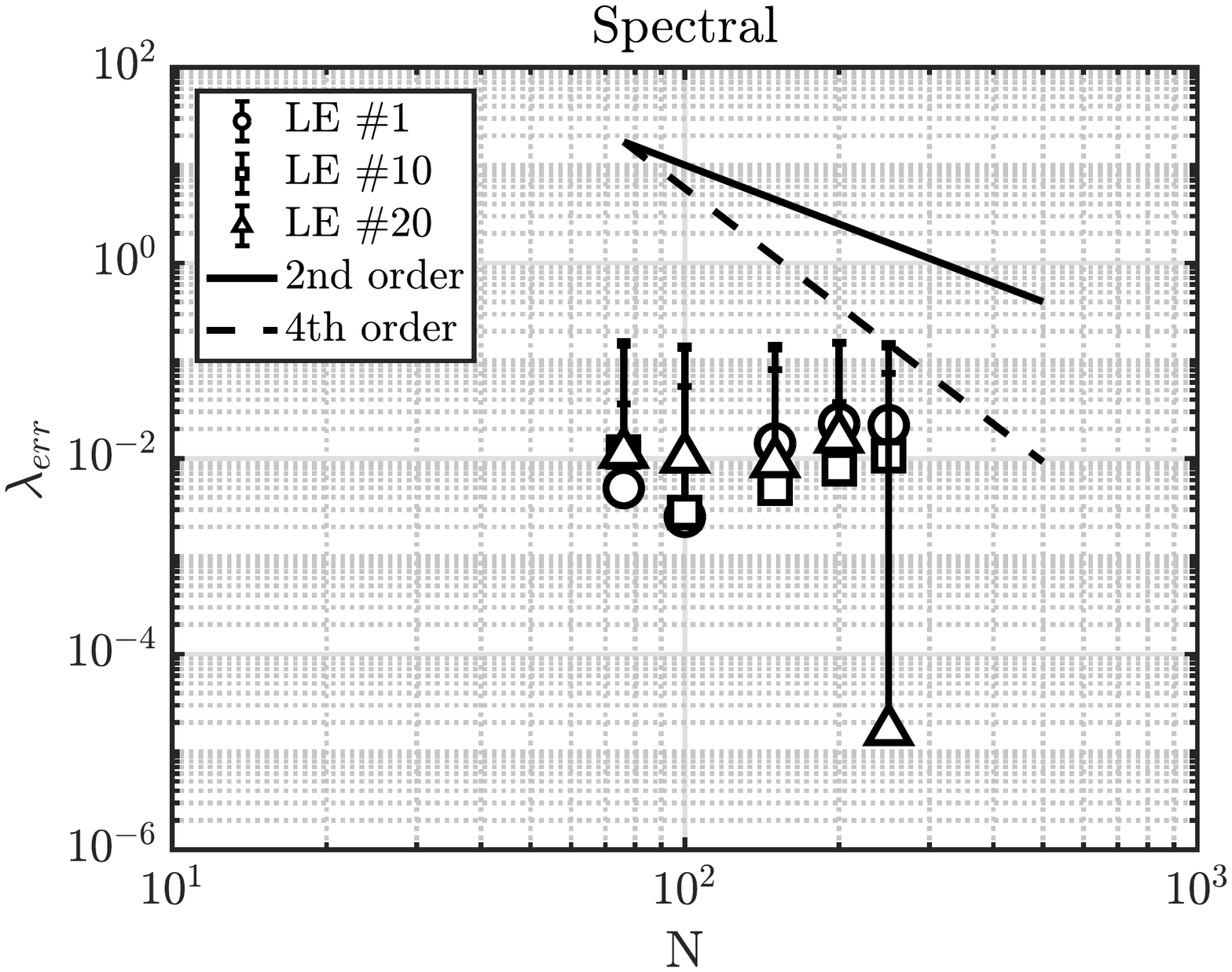}
\caption{Convergence of the $1^{st}$, $10^{th}$, $20^{th}$ LE with discretization level. Left: $2^{nd}$ order. Middle: $4^{th}$ order. Right: spectral method.}
\label{fig:KSbad}
\end{figure}

Figure~\ref{fig:KSgood} shows the convergence for the $35^{th}$, $40^{th}$ and $45^{th}$ LEs. Here the errors reported are absolute errors. It is seen that the spatial convergence of this set of LEs is directly related to the discretization order, with very high convergence rates (albeit not exponential) for the LEs computed using spectral derivatives. These results are consistent with the planar Couette flow findings (Sec.~\ref{sec:couette}). However, it is possible to find other LEs that do not conform to this trend. For instance, Fig.~\ref{fig:KSbad} show spatial convergence plots for $1^{st}$, $10^{th}$ and $20^{th}$ LEs.

\begin{figure}
\center
\includegraphics[width=0.45\textwidth,trim={0cm 0cm 0cm 0cm},clip]{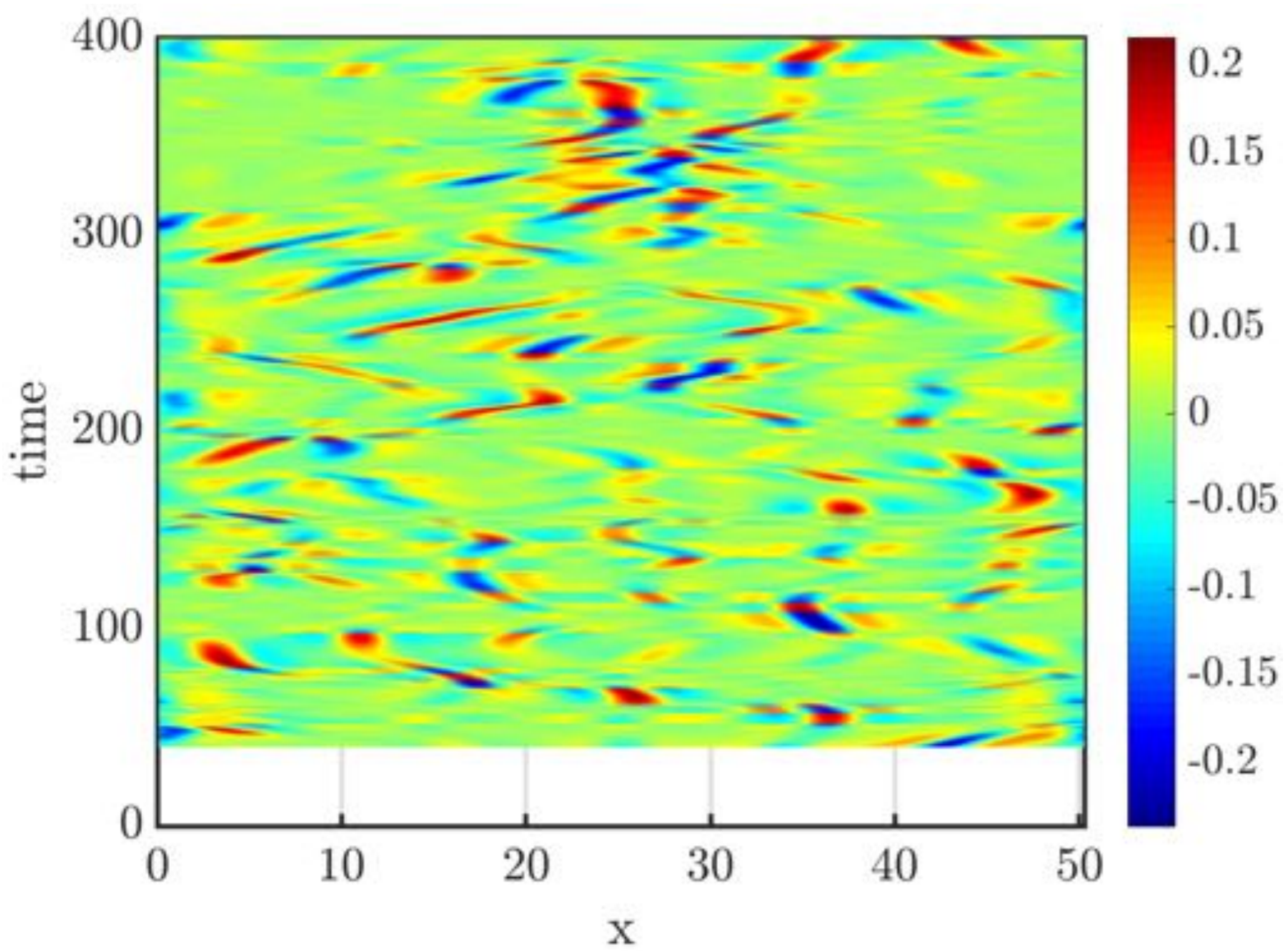}
\includegraphics[width=0.45\textwidth,trim={0cm 0cm 0cm 0cm},clip]{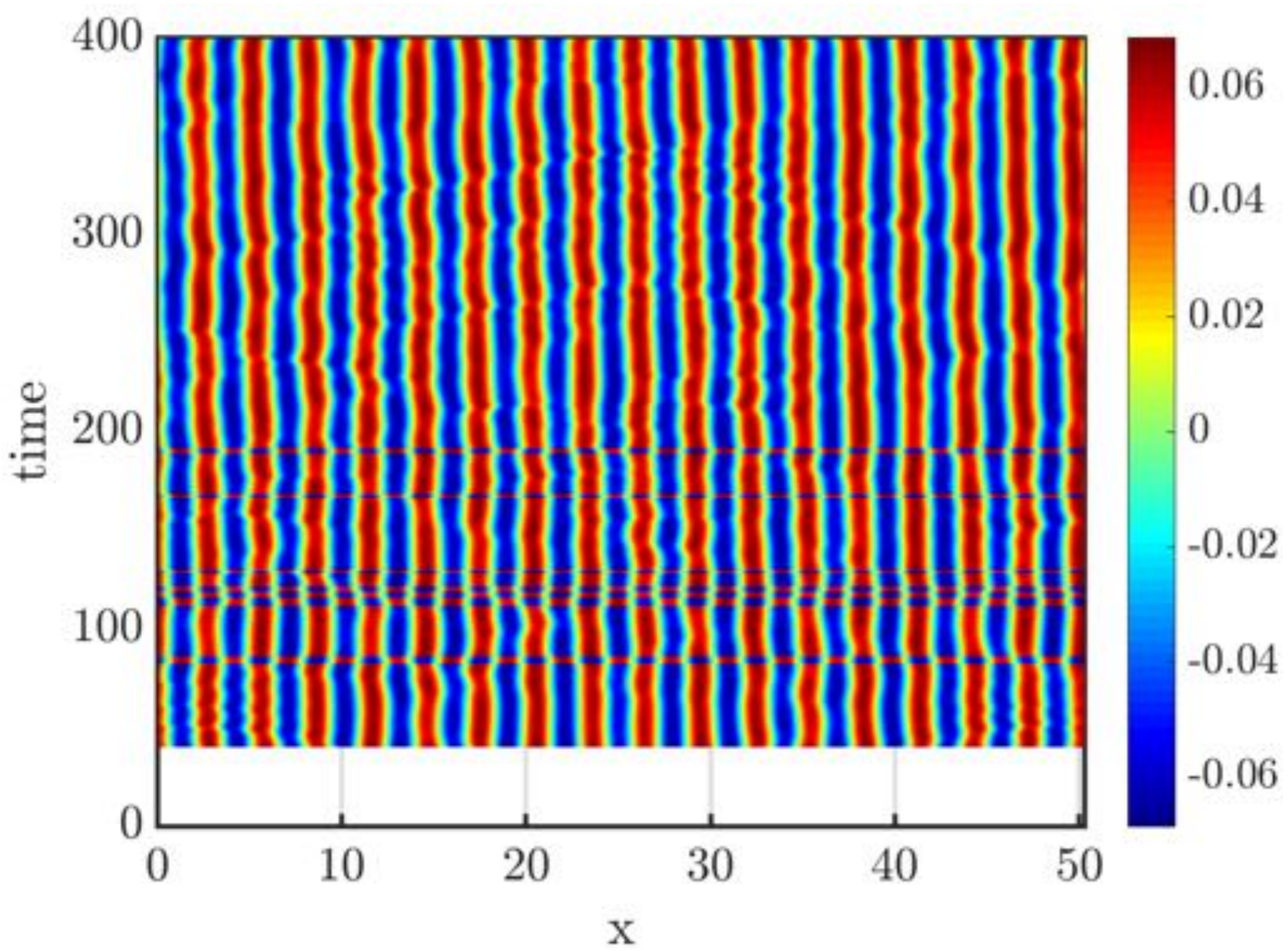}
\caption{Time evolution of the 1st LV (left) and 35th LV (right) of the KS equation computed with 500 modes.}
\label{fig:KS_LV}
\end{figure}

To gain further insight, the space-time plot for two representative LVs ($1^{st}$ and $35^{th}$) are plotted in Fig.~\ref{fig:KS_LV}. Interestingly, the $35^{th}$ LV shows very little variation with time, and very organized structure with nearly periodic wave patterns. The first LV, on the other hand, shows a more chaotic structure that is similar to the KS solution itself. We postulate that the difference in convergence properties is due to the associated smoothness of these solutions. The $35^{th}$ LV (and the other convergent vectors) exhibits a regular spatial structure that is smooth with respect to the grid spacing. On the other hand, the non-converging LVs exhibit a broadband spectrum of fluctuations which may amplify the dispersion errors. It is also worth noting that even when the LVs do not converge with grid spacing, the absolute errors are relatively low (Fig.~\ref{fig:KSbad}).

This study indicates that the LS could have LE-dependent convergence rates, which should be treated as a note of caution when obtaining the spectrum. In particular, the lack of spatial convergence for the first LE is noteworthy.

\subsection{Turbulent channel flow}

Chaotic flow in a rectangular channel is an important canonical configuration that is often used for deriving turbulence theories. Further, multiple studies \cite{keefe,inubushi} have utilized this configuration for studying the properties the Lyapunov spectrum. For this reason, the flow conditions used here replicate the study of Keefe et al.~\cite{keefe} corresponding to $Re_{\tau}=80$ and a domain size of $1.6 \pi \times 2 \times 1.6 \pi$. This configuration is simulated using the low-Mach number algorithm discussed in Sec.~\ref{sec:algorithm}. 

\begin{table}[ht!]
\centering
 \begin{tabular}{|c | c | c | c | c | c |} 
 \hline
 Resolution ($x \times y \times z$) & $\frac{k_w \Delta t u_{\tau}}{L}$ & $\frac{k_s \Delta t u_{\tau}}{L}$ & $m$ & $\frac{N_s k_s \Delta t u_{\tau}}{L}$ & $\Delta t$ \\
 \hline \hline
 $16 \times 32 \times 16$ & 0.0038 & 0.065 & 110 & 236.77 &  0.01s   \\ 
 \hline
 $32 \times 64 \times 32$ & 0.0038 & 0.065 & 110 & 109.29 & 0.01s \\
 \hline
 $64 \times 128 \times 64$ & 0.0038 & 0.065 & 27 & 193.5 & 0.01s  \\
 \hline
\end{tabular}
\caption{Details of the grid size and parameters for the LS algorithm. $u_{\tau}$ denotes the wall shear velocity, and $L$ is the channel width.}
\label{tab:turbChan}
\end{table}

To understand the role of spatial discretization on the LS, simulations with different grid resolutions are carried out. The mesh details along with the parameters for the LS algorithm are provided in Tab.~\ref{tab:turbChan}. Note that for all the resolutions, the time-step is held fixed at $1 \times 10^{-2}$s which allows to isolate the effects of spatial discretization. The spatial derivatives were obtained using a second-order approach, while the time-stepping also utilized a second-order accurate scheme.

\begin{figure}[h]
\center
\includegraphics[width=0.43\textwidth,trim={0cm 0cm 0cm 0cm},clip]{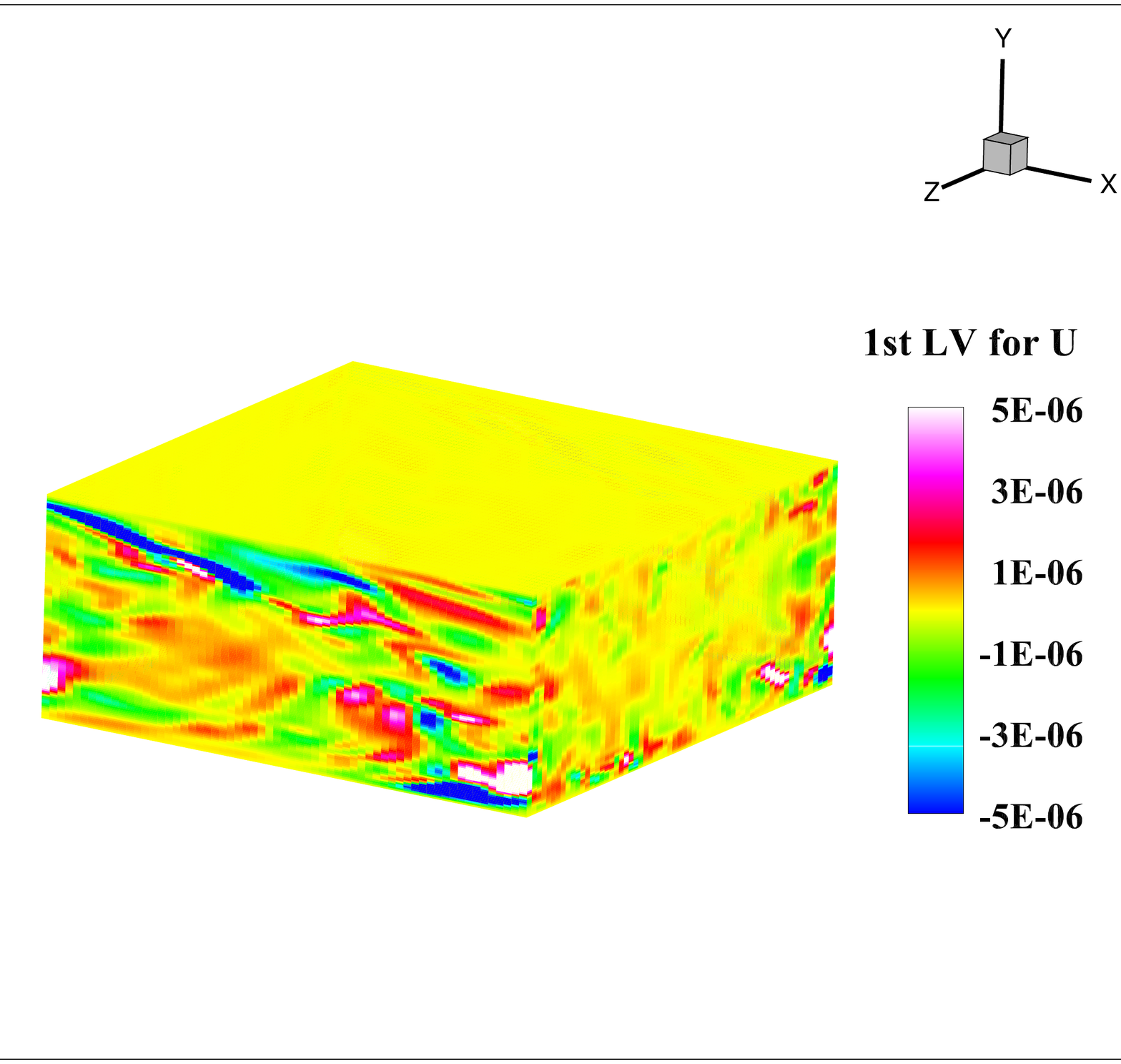}
\includegraphics[width=0.4\textwidth,trim={0cm 0cm 0cm 0cm},clip]{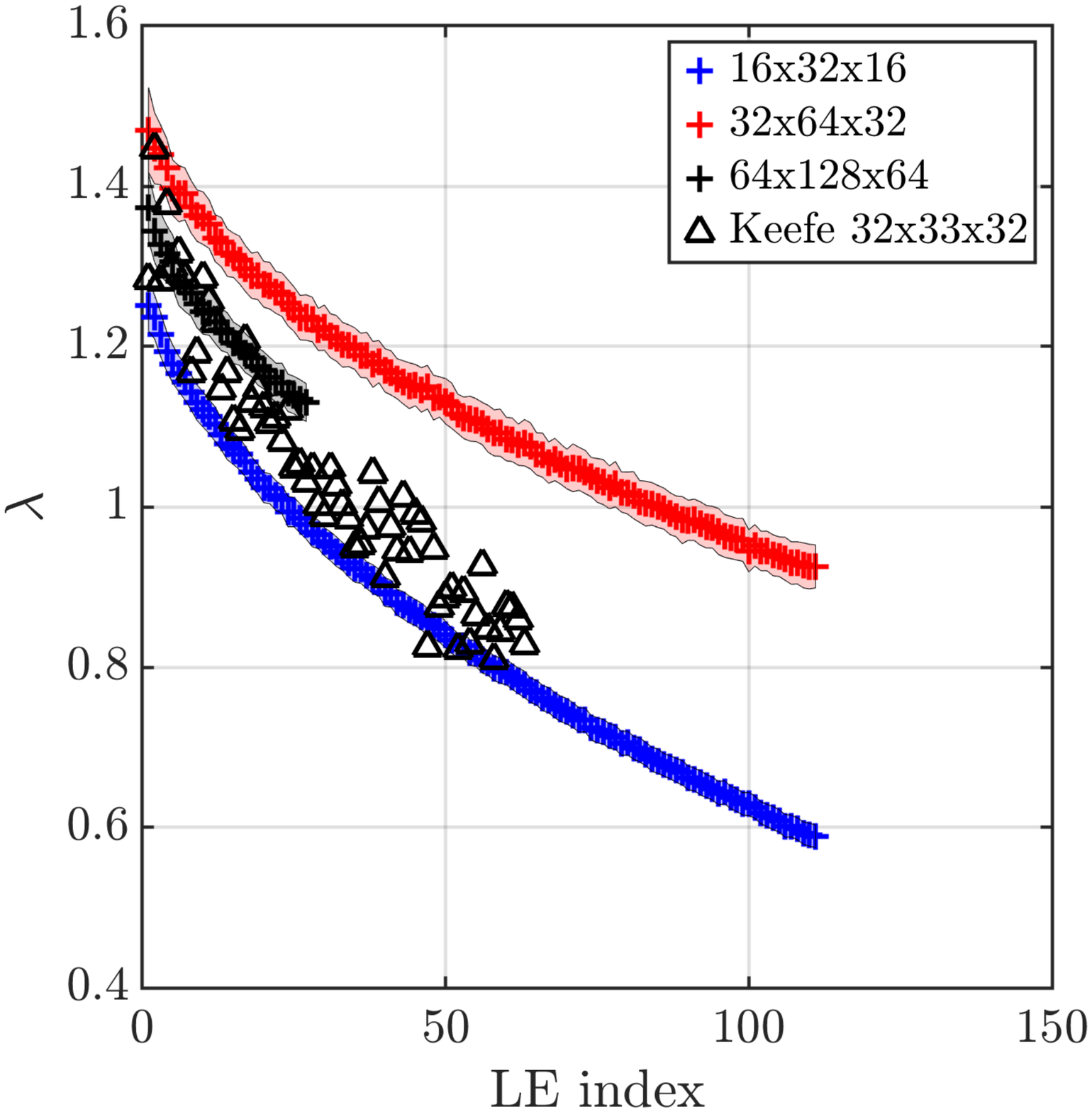}
\caption{Left: instantaneous contour of the first Lyapunov vector projected in the x component of velocity for the $64 \times 128 \times 64$ simulation. Right: Lyapunov spectrum obtained with different grids, overlayed with sampling uncertainty (shades). Data from Keefe et al.~\cite{keefe} is added for comparison.} 
\label{fig:specTurbChan}
\end{figure}

Figure~\ref{fig:specTurbChan} shows the set of LEs computed for all three resolutions. The results are shown along with the sampling uncertainty (shaded area) due to the limited time-average of the exponents. This uncertainty is estimated similarly to the study of the KS equation discussed in Sec.~\ref{sec:ks}, using the time series of each LE \cite{uncertaintyDNS_oliver}. The LEs were averaged for a long enough time ($\frac{N_s k_s \Delta t u_{\tau}}{L}$) so as to ensure that sampling uncertainty does not affect the interpretation of the results. Additionally, the results obtained by Keefe et al.~\cite{keefe} using a spectral approach and a grid resolution of $32 \times 33 \times 32$ is shown. Overall, the LE obtained are close to the results of Keefe et al.~\cite{keefe}. Small deviations can be observed in terms of the exponents, for instance in the slope of the LS. Interestingly, only the coarsest and finest grid calculations match the slope obtained by Keefe et al.~\cite{keefe} while the medium resolution overestimated the LE. Further investigation showed that for the coarsest and finest grids, the turbulent channel operated at a near constant Re$_\tau$ of 80, while the medium resolution case maintained Re$_\tau$ of 83.6. This could explain the apparent lack of spatial convergence for the spectrum. This observation is important as it highlights that the convergence of the LS is dependent on the convergence of the underlying flow field. Similar correlations were observed by Fernandez et al.~\cite{fernandez} for a different flow configuration.

\subsection{Homogeneous Isotropic Turbulence (HIT)}

\begin{table}
\centering
 \begin{tabular}{|c | c | c | c | c | c | c |} 
 \hline
 Resolution & $ Re_{\lambda} $ & $\frac{k_w \Delta t}{\tau_{eddy}} $ & $\frac{k_s \Delta t}{\tau_{eddy}}$ & $m$ & $\frac{N_s k_s \Delta t}{\tau_{eddy}}$ & $\Delta t$ \\
 
 \hline \hline
 $32^3$ & 12.9 & 0.038 & 1.06 & 100 & 6707 &  0.02s   \\ 
 \hline
 $64^3$ & 12.9 & 0.038 & 1.06 & 55 & 7028 & 0.02s \\
 \hline
 $128^3$ & 12.9 & 0.038 & 1.06 & 55 & 1013 & 0.02s  \\
 \hline
 $256^3$ & 12.9 & 0.038 & 1.06 & 55 & 446 & 0.02s  \\
 \hline
\end{tabular}
\caption{Mesh resolution and simulation parameters for the LS computation for HIT, where $\tau_{eddy}$ denotes the eddy turnover time.}
\label{tab:hit}
\end{table}

Another typical theoretical configuration for turbulence studies is the HIT flow in a $2\pi^3$ periodic box. Previously, the largest LE (i.e., the first LE) has been computed by Mohan et al.~\cite{mohan2017scaling}, who found that time-scale associated with this LE decreases faster than the Kolmogorov time scale. Here, a partial spectrum of LEs is computed. The HIT simulation is conducted with the low Mach number with 2$^{nd}$ order spatial and temporal accuracy. Other relevant simulation details are summarized in Table.~\ref{tab:hit}. The grid resolutions are chosen such that the smallest turbulence length scale is resolved by the coarsest grid. The turbulence is sustained using a linear forcing method \cite{Rosales:2005fy} with a forcing coefficient of $0.1$. The resulting Taylor microscale Reynolds number is $Re_{\lambda} = 12.94$ for all simulations as indicated in Tab.~\ref{tab:hit}. For all the spatial resolutions the time-step was held constant at $0.02$s, which is based on the CFL condition for the finest mesh resolution.

\begin{figure}
\center
\includegraphics[width=0.45\textwidth,trim={0cm 0cm 0cm 0cm},clip]{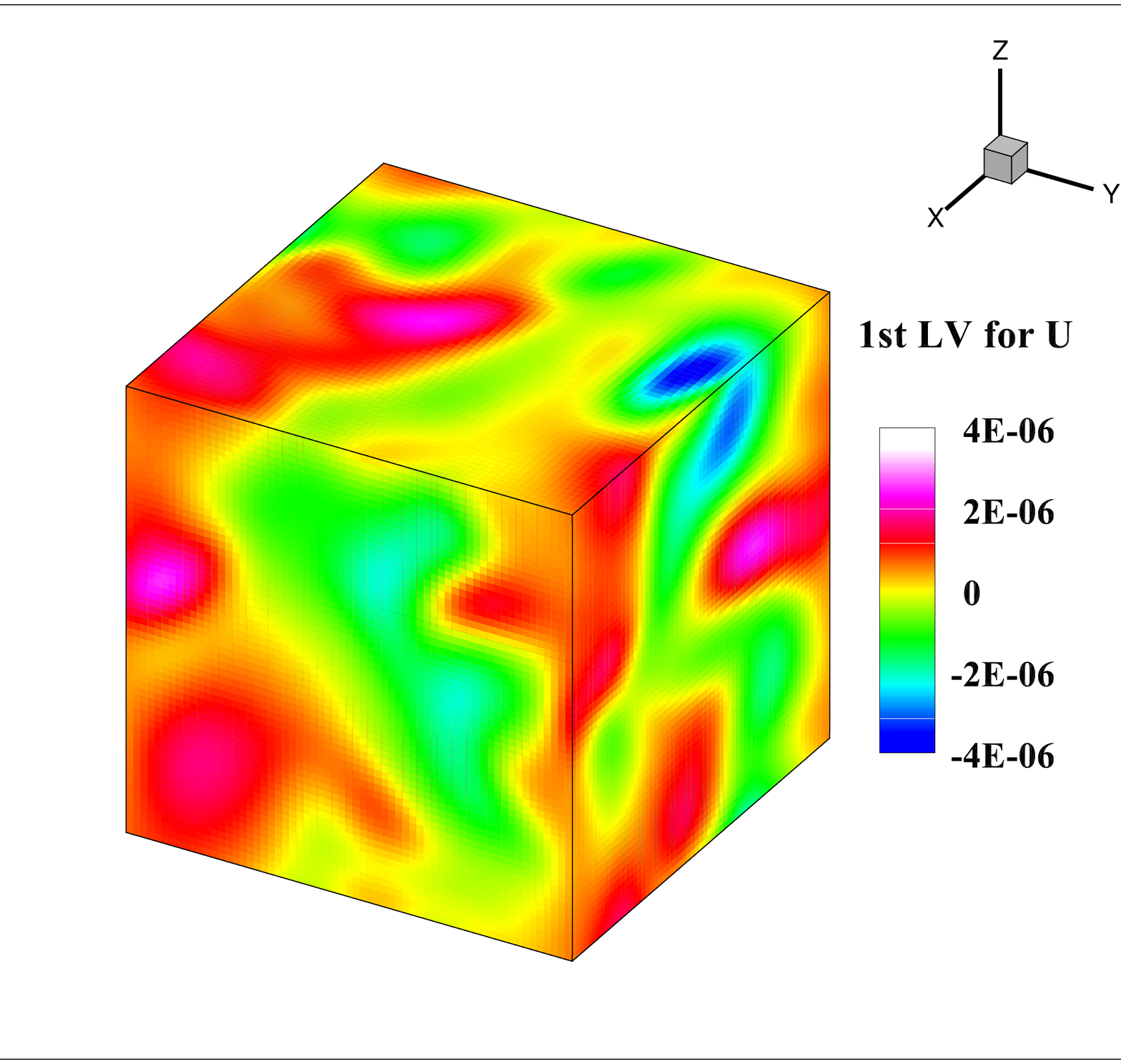}
\includegraphics[width=0.42\textwidth,trim={0cm 0cm 0cm 0cm},clip]{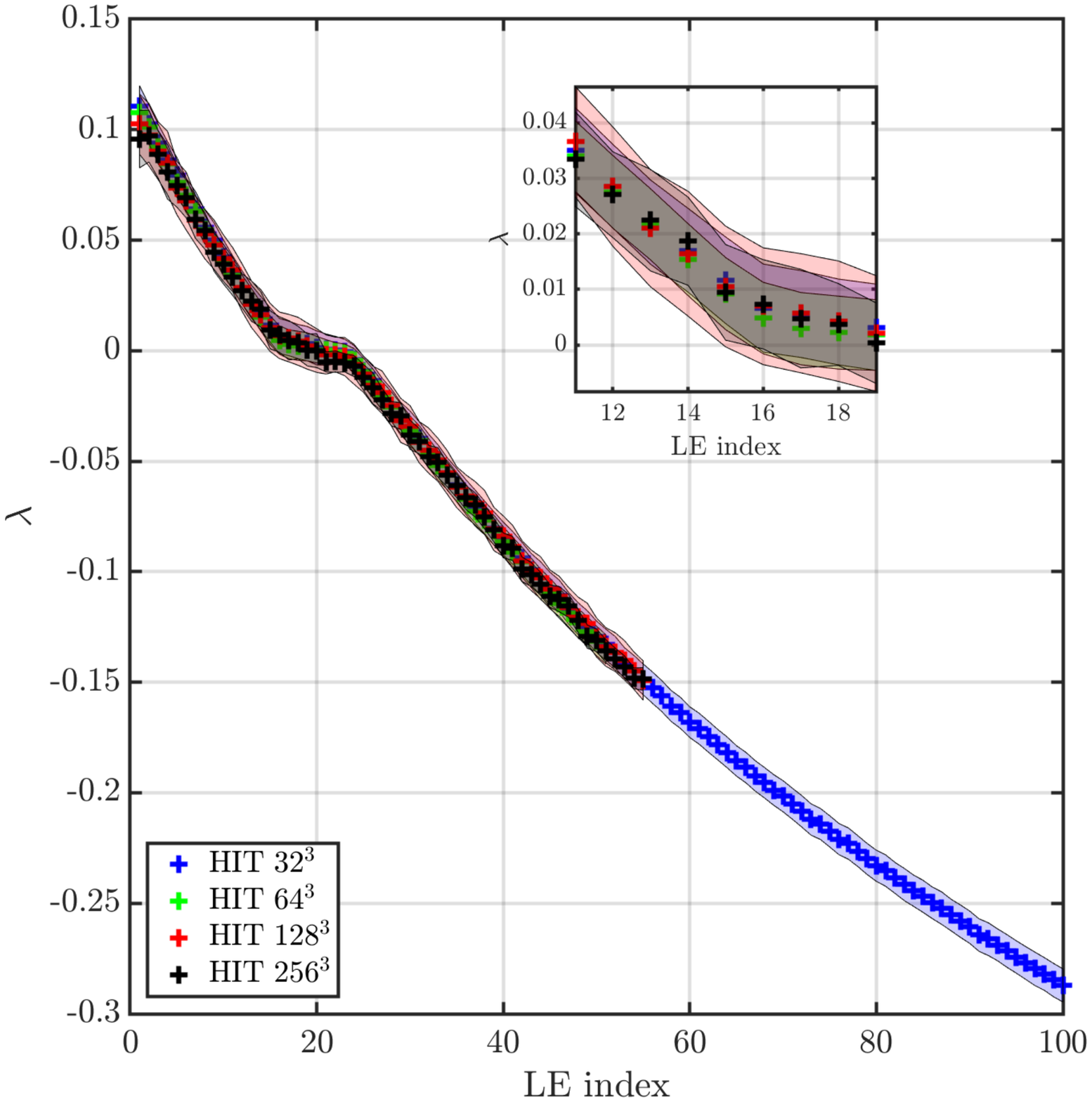}
\caption{Left: instantaneous contour of the first Lyapunov vector projected in the x component of velocity for the $64^3$ homogeneous isotropic turbulent simulation. Right: Lyapunov spectrum obtained with different spatial resolutions. Right inset: zoom on the LE between the 12$^{th}$ and the 18$^{th}$ index.}
\label{fig:hitSpec}
\end{figure}

Figure~\ref{fig:hitSpec} shows the spectrum obtained with all four resolutions along with the standard deviation in the estimation of the average LEs. Since the Reynolds number is low, the spectrum appears converged for all resolutions considered. Further, three distinctive parts of the spectrum can be observed:  1) the chaotic part characterized by positive Lyapunov exponents, for indices ranging from 1-18,  2) a ``knee-like" structure around index $20$ with LEs straddling zero value, which is characteristic of perturbations that do not expand or contract on average, and 3) a dissipation range starting from index $22$. Interestingly, the slope of the spectrum in the chaotic and the dissipation ranges are roughly identical. The knee-like behavior has been observed in other dynamical systems with multiple exponents that are close to zero \cite{van2009lyapunov,karimi2010extensive}.

\begin{figure}
\center
\includegraphics[width=0.4\textwidth,trim={0cm 0cm 0cm 0cm},clip]{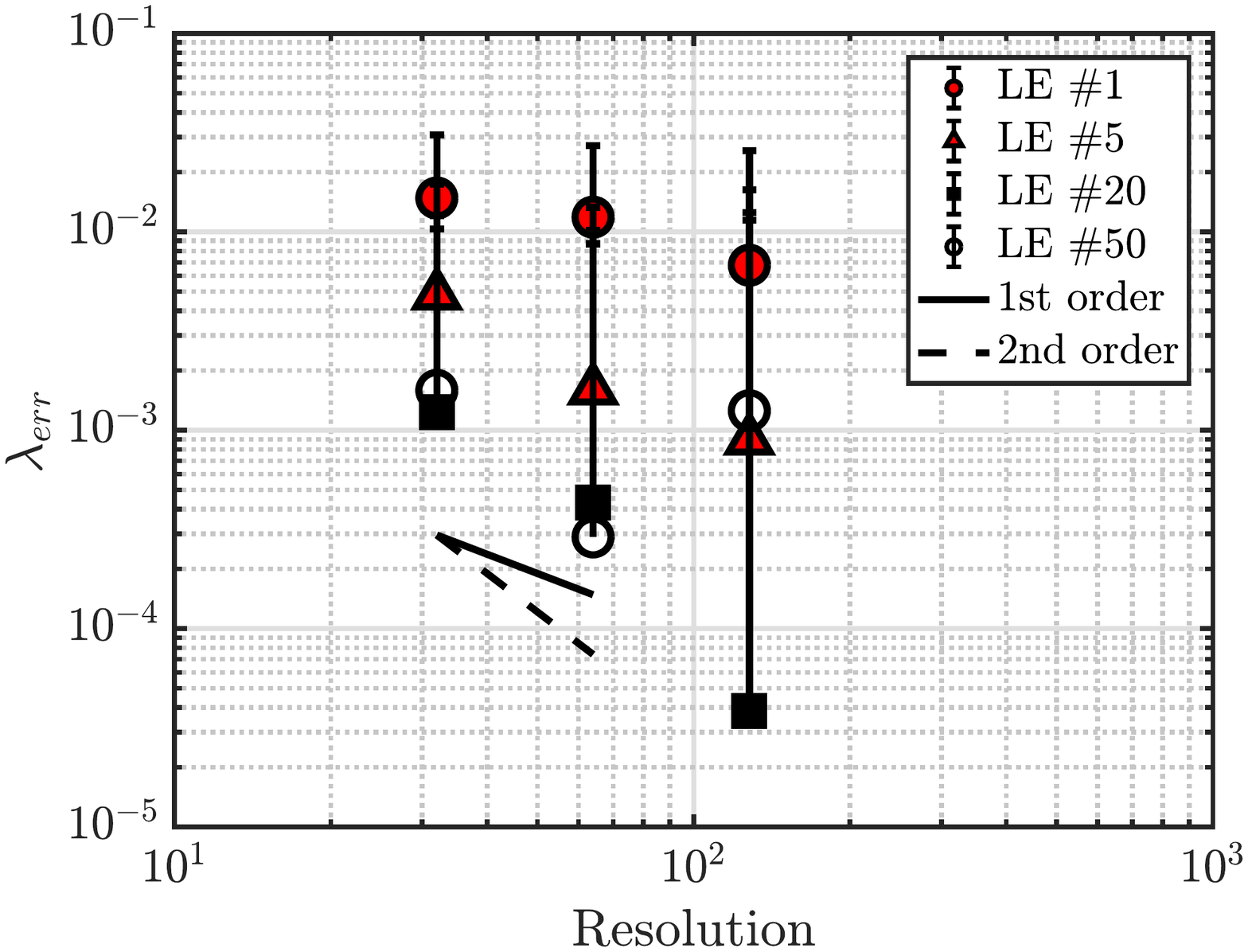}
\includegraphics[width=0.4\textwidth,trim={0cm 0cm 0cm 0cm},clip]{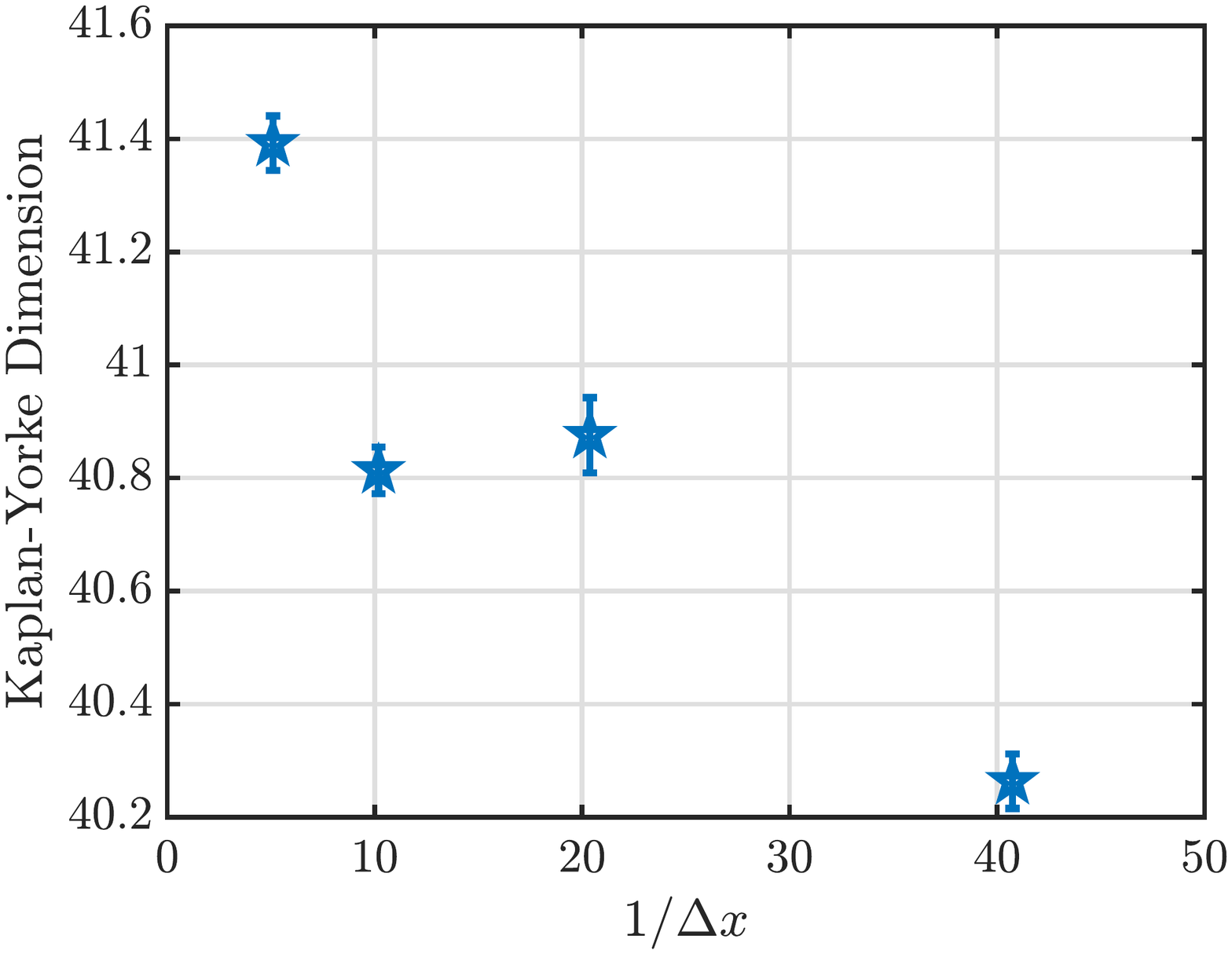}
\caption{Left: dependence of the LEs of the HIT simulation with spatial discretization. Right: dependence of the Kaplan-Yorke dimension of the HIT simulation with the spatial discretization}
\label{fig:convergenceLEDim}
\end{figure}

Using the results of the finest-mesh simulation as the ``true" spectrum, the convergence properties of various exponents can be examined (Fig.~\ref{fig:convergenceLEDim}). It is seen that there is large uncertainty in the estimated errors, primarily due to large fluctuations in the FTLEs. Hence, it is difficult to draw conclusive evidence from these results. Nevertheless, focusing on the mean LE, there are differences in the convergence rates for different regions of the spectrum. In the chaotic regime, a first order convergence is observed, while in the knee and dissipative regimes, second order convergence is obtained. These findings are consistent with results from the KS study, where also LE-dependent convergence rates were obtained. 

Based on the LS, the Kaplan-Yorke dimension of the attractor can be estimated as \cite{kaplan1979chaotic}:
\[
  D_{KY} = j + \frac{\sum_{i=1}^{i=j} \lambda_i}{\abs{\lambda_{j+1}}}~~\text{where}~~ \sum_{i=1}^{i=j} \lambda_i \geq  0~~\text{and} ~~ \sum_{i=1}^{i=j+1} \lambda_i \leq 0.
\]
The dimension estimates obtained for all cases are shown in Fig.~\ref{fig:convergenceLEDim}. The variations in dimension is relatively small as the grid size is altered, and is roughly 2\% of the mean value. First, it appears that the dimension obtained for each resolution vary from each other by as little as $2\%$. However, the convergence of the dimension with respect to the grid resolution is not directly evident, even if only the upper-bound of the dimension is considered. This suggests that the flow field has been slightly affected by the increased resolution even at DNS resolution. Interestingly, the dimension decreased with the increasing resolution, while an opposite trend  was observed by Fernandez et al.~\cite{fernandez}.

\section{Conclusions}

The computation of the Lyapunov spectrum using low Mach number solvers introduces several challenges that are addressed in this work. First, the definition of state-space depends on the numerical procedure adopted by the solver. Failure to include a variable in the state-vector can lead to errors in the estimation of the Lyapunov exponents. Second, the non-mass-conservative orthogonalized state-vectors lead to large variations in the initial step of the calculation of the Lyapunov exponents. In particular, such perturbations will push the system out of the attractor, with the calculated exponents dominated by the return trajectory of the system towards the attractor. A procedure that alleviates this problem by estimating the time required to reach the attractor has been introduced. Using these algorithms, the issue of spatial and temporal convergence of Lyapunov spectrum was studied using a number of canonical flow problems. The Orr-Sommerfeld (OS) equations were used to derive an analytical relation between the OS eigenvalues and Lyapunov exponents, which provides a highly accurate approach for verifying the numerical algorithms. A sequence of canonical configurations showed interesting results. First, the spatial and temporal convergence are highly problem dependent, with some flow configurations producing convergence rates that are directly related to the truncation errors in the numerical algorithms used. In other cases, regardless of the spatial resolution, convergence of exponents could not be obtained. Second, within a particular configuration, different Lyapunov exponents might exhibit different convergence rates. In the Kuramoto-Sivashinsky (KS) equation as well as homogeneous isotropic turbulence (HIT) cases, it was found that select exponents and their associated vectors converged at the rate of the underlying discretization order, while other exponents showed non-convergent behavior.

Finally, the Lyapunov spectrum found in this work for the different canonical flow problems deserves additional studies. For instance, the HIT spectrum showed three distinctive regions, which to our knowledge, is the first such observation for this problem. The KS equation shows that certain Lyapunov vectors are highly localized in spectral space, while other vectors show turbulence-like spectrum. These results show an extraordinary richness in the chaotic behavior of these systems that is yet to be explored. These topics will be pursued in future studies.

\section{Acknowledgements}

This work was financially supported by an AFOSR research grant (FA9550-15-1-0378) with Dr.\,Chiping Li as program manager. The authors thank NASA HECC for generous allocation of computing time on NASA Pleiades machine.

\bibliographystyle{elsarticle-num}

\end{document}